\documentclass[iop,numberedappendix,appendixfloats,twocolappendix,revtex4]{emulateapj}

\usepackage{amsmath, latexsym, color, verbatim}
\usepackage{graphicx,epsfig, natbib, epstopdf}
\usepackage{pdflscape}
\usepackage{ulem}

\lefthead{Law et al. 2020}
\righthead{SDSS-IV MaNGA: Modeling the Spectral LSF}

\slugcomment{DRAFT: \today}

\begin{document}

\newcommand{\msun}{\ensuremath{\rm M_\odot}}
\newcommand{\msunyr}{\ensuremath{\rm M_{\odot}\;{\rm yr}^{-1}}}
\newcommand{\Ha}{\ensuremath{\rm H\alpha}}
\newcommand{\Hb}{\ensuremath{\rm H\beta}}
\newcommand{\lya}{\ensuremath{\rm Ly\alpha}}
\newcommand{\Ntwo}{[\ion{N}{2}]}
\newcommand{\kms}{{km$~\!$s$^{-1}$}}
\newcommand{\ztwo}{\ensuremath{z\sim2}}
\newcommand{\zthree}{\ensuremath{z\sim3}}
\newcommand{\feh}{\textrm{[Fe/H]}}
\newcommand{\afeh}{\textrm{[$\alpha$/Fe]}}
\newcommand{\nifeh}{\textrm{[Ni/Fe]}}

\newcommand{\sigha}{\ensuremath{\sigma_{\Ha}}}

\newcommand{\oone}{\textrm{O\,{\sc i}}}
\newcommand{\ofour}{\textrm{[O\,{\sc iiii}]}}
\newcommand{\othree}{\textrm{[O\,{\sc iii}]}}
\newcommand{\otwo}{\textrm{[O\,{\sc ii}]}}
\newcommand{\ntwo}{\textrm{[N\,{\sc ii}]}}
\newcommand{\stwo}{\textrm{[S\,{\sc ii}]}}
\newcommand{\sthree}{\textrm{[S\,{\sc iii}]}}

\newcommand{\redtxt}[1]{\textcolor{red}{#1}}

\title{SDSS-IV MaNGA: Modeling the Spectral Line Spread Function to Sub-Percent Accuracy}

\author{David R.~Law\altaffilmark{1}, 
Kyle B.~Westfall\altaffilmark{2},
Matthew A.~Bershady\altaffilmark{3,4,5}, 
Michele Cappellari\altaffilmark{6},
Renbin Yan\altaffilmark{7}, 
Francesco Belfiore\altaffilmark{8}, 
Dmitry Bizyaev\altaffilmark{9}, 
Joel R.~Brownstein\altaffilmark{10},
Yanping Chen\altaffilmark{11},
Brian Cherinka\altaffilmark{1}, 
Niv Drory\altaffilmark{12}
Daniel Lazarz\altaffilmark{7}, 
Shravan Shetty\altaffilmark{3,13}
}

\altaffiltext{1}{Space Telescope Science Institute, 3700 San Martin Drive, Baltimore, MD 21218, USA; dlaw@stsci.edu}
\altaffiltext{2}{University of California Observatories, University of California, Santa Cruz, 1156 High St., Santa Cruz, CA 95064, USA.}
\altaffiltext{3}{University of Wisconsin - Madison, Department of Astronomy, 475 N. Charter Street, Madison, WI 53706-1582, USA.}
\altaffiltext{4}{South African Astronomical Observatory, PO Box 9, Observatory 7935, Cape Town, South Africa.}
\altaffiltext{5}{Department of Astronomy, University of Cape Town, Private Bag X3, Rondebosch 7701, South Africa.}
\altaffiltext{6}{Sub-department of Astrophysics, Department of Physics, University of Oxford, Denys Wilkinson Building, Keble Road, Oxford OX1 3RH, UK.}
\altaffiltext{7}{Department of Physics and Astronomy, University of Kentucky, 505 Rose Street, Lexington, KY 40506-0057, USA.}
\altaffiltext{8}{INAF -- Osservatorio Astrofisico di Arcetri, Largo E. Fermi 5, I-50157, Firenze, Italy.}
\altaffiltext{9}{Apache Point Observatory, P.O. Box 59, Sunspot, NM 88349, USA.}
\altaffiltext{10}{University of Utah, Department of Physics and Astronomy, 115 S. 1400 E., Salt Lake City, UT 84112, USA}
\altaffiltext{11}{New York University Abu Dhabi, P.O. Box 129188, Abu Dhabi, UAE.}
\altaffiltext{12}{McDonald Observatory, The University of Texas at Austin, 2515 Speedway, Stop C1402, Austin, TX 78712, USA.}
\altaffiltext{13}{Kavli Institute for Astronomy and Astrophysics, Peking Uni- versity, Beijing 100871, China.}

\begin{abstract}

The SDSS-IV Mapping Nearby Galaxies at APO (MaNGA) program has been operating from 2014-2020, and has now observed a sample
of $9,269$ galaxies in the low redshift universe ($z \sim 0.05$) with integral-field spectroscopy.
With rest-optical  ($\lambda\lambda 0.36 - 1.0 \micron$) spectral resolution $R \sim 2000$ the instrumental spectral line-spread function (LSF) typically has
$1\sigma$ width of about 70 \kms, which poses a challenge for the study of the typically 20-30 \kms\
velocity dispersion of the ionized gas in present-day disk galaxies.
In this contribution, we present a major revision of the MaNGA data pipeline architecture,
focusing particularly on
a variety of factors impacting the effective LSF (e.g., undersampling, spectral rectification, and data cube construction).
Through comparison with  external assessments of the MaNGA data provided by substantially 
higher-resolution $R \sim 10,000$ instruments we 
demonstrate that the revised MPL-10 pipeline
measures the instrumental line spread function sufficiently accurately ($\leq$ 0.6\% systematic, 2\% random around the wavelength of \Ha)
that it enables reliable measurements of astrophysical velocity dispersions 
$\sigma_{\Ha} \sim 20$ \kms\ for 
spaxels with emission lines detected at SNR $> 50$.
Velocity dispersions derived from \otwo, \Hb, \othree, \ntwo, and \stwo\ are consistent with those derived from \Ha\ to within about
2\% at $\sigma_{\Ha} > 30$ km s$^{-1}$.
Although the impact of these changes to
the estimated LSF will be minimal at velocity dispersions greater than about 100 \kms, 
scientific results from previous data releases that are based on dispersions far
below the instrumental resolution should be reevaulated.

\end{abstract}



\section{Introduction}

The design of astrophysical instruments is always a trade-off between various competing factors.  With limited detector real estate, there is an inherent
tension between (i) the number of spectra that can be observed, (ii) the wavelength range that they can cover, and (iii) the spectral resolution (or the 
effective information content per wavelength).

The first is obviously attractive; the more spectra that can be observed simultaneously the faster any survey can be completed, or the larger its
eventual sample of objects.  The second is similarly obvious; the rest-optical and NIR wavelength range is replete with a wealth of spectral features
encoding information about the kinematics, stellar populations, chemical abundances, and sources of ionizing radiation.
For the SDSS-IV MaNGA survey for instance \citep[][]{bundy15,blanton17}, the design of the instrument has allowed the survey to observe a sample of
10,000 galaxies with IFU spectroscopy (roughly a factor $\sim 10$ larger than previous such surveys),
at the same time as spanning a wide
and contiguous wavelength range from $3600 - 10,300$ \AA.  This wavelength range crucially includes classic strong-line emission 
features from \otwo\ $\lambda3727$ to \sthree\ $\lambda 9531$ that can characterize the mechanisms of ongoing star formation,
stellar absorption features such as the Mg triplet at 5170 \AA\ and the Ca triplet at 8550 \AA\ 
that characterize the evolved population, and faint indices such as NaI at 8120 \AA\ and FeH at 9916 \AA\ that are sensitive to the initial mass function
\citep[e.g.,][]{cd12,parikh18}.

At the same time, high spectral resolution is critical to both separate spectral features that are close together in wavelength (e.g., the [\ion{O}{2}] 3727 doublet) and to study
the range of velocities present in gas or stellar populations along a given line of sight.
Given the necessity of high spectral resolution in studies of galaxy
kinematics, IFU surveys such as SAMI \citep[][]{croom12,allen15} have opted to trade spectral coverage for higher spectral resolution
$R \sim 4500$ around key diagnostic features such as \Ha.
In contrast, at the spectral resolution of MaNGA ($R \sim 2000$) it is much more difficult to extract kinematic data as the astrophysical
ionized gas velocity dispersions are typically $\sigha \sim 20-30$ \kms\ for main sequence star forming galaxies. These astrophysical dispersions are dwarfed by the 
$\sim 70$ \kms\ line widths produced by the instrumental line spread function (LSF; i.e. the projection of the detector point spread function
onto the spectral axis).\footnote{For reference, a 20 \kms\ broadening of an $R=2000$ LSF results in a 5\% increase in the width of the observed line, whereas this same broadening results in a 22\% increase at $R=4500$.}
Complicating matters further, MaNGA is a critically-sampled ($\sim2$ pixels per FWHM) fiber-based instrument that feeds a pair of two-arm, Cassegrain-mounted spectrographs subject to varying gravitational flexure.  As we show here, this leads to an instrumental LSF that is highly variable, spatially (from fiber to fiber), spectrally (both between arms and within each arm), and temporally.

\begin{figure}
\begin{center}
\includegraphics[width=0.9\columnwidth]{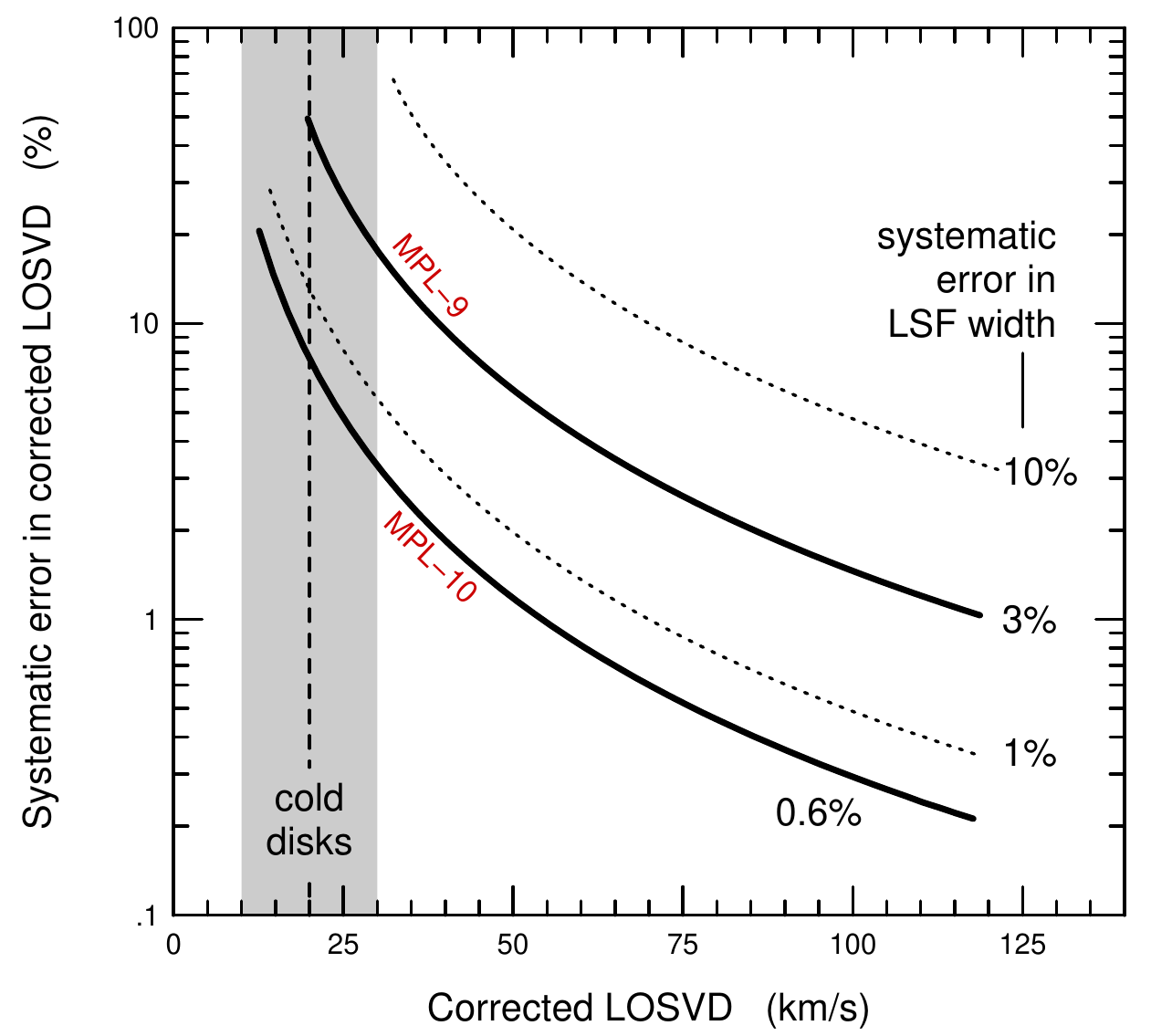}
\end{center}
\caption{Systematic error in  derived $\sigma_{\Ha}$ as a function of $\sigma_{\Ha}$ for various assumed errors in the MaNGA instrumental LSF. This is estimated analytically from the standard assumption that the instrumental and astrophysical dispersion add in quadrature.  In order to achieve better than 10\% accuracy in the regime of cold gas disks the LSF must be known to better than 1\%.
}
\label{requirement.fig}
\end{figure}

To reliably recover astrophysical line widths on the order of $20$ \kms, this instrumental contribution must be modeled in exquisite detail
and accurately removed from the measured line widths (emphasizing that each of these requirements present unique challenges).
In this contribution, we present a major update to the original MaNGA data pipeline  \citep{law16} 
and demonstrate that the most recent MPL-10 survey data products 
meet (and indeed exceed) the 1\% level of precision in the LSF necessary to study such astrophysical signals (see Figure \ref{requirement.fig}). 

This paper is organized as follows:
In \S \ref{obs.sec} we provide an overview of the MaNGA instrument and observing program,  and highlight major changes that have been made
to the Data Reduction Pipeline (DRP) since the initial description by \citet{law16}.
In \S \ref{drp.sec} we discuss the revised derivation of the instrumental
LSF in the latest version of the MaNGA DRP (MPL-10; see
Table \ref{versions.table}), noting differences from earlier calibrations and publicly-released data products.
We describe changes that have been made to the MaNGA Data Analysis
Pipeline (DAP) since the earlier description by \citet{westfall19}
in \S \ref{dap.sec} and use internal tests to demonstrate the reliability of our estimates of the astrophysical
gas and stellar velocity dispersions.
We extend this discussion to consider the effects of beam smearing in \S \ref{beamsmear.sec}.
Finally, in \S \ref{external.sec} we compare the MaNGA data directly against independent external calibrators at much higher spectral
resolution (up to $R \sim 11,000$) and demonstrate that the instrumental LSF estimates provided by the pipeline are accurate
to better than 1\% around the wavelength of \Ha.
We summarize our conclusions in \S \ref{summary.sec}.

Throughout our analysis we adopt a \citet{chabrier03} stellar initial mass function and a 
$\Lambda$CDM cosmology in which $H_0 = 70$ km s$^{-1}$ Mpc$^{-1}$, $\Omega_m = 0.27$, and$\Omega_{\Lambda} = 0.73$.


\section{Observational Data}
\label{obs.sec}

The MaNGA hardware design is described in detail by \citet{drory15}, and consists of 1423 optical fibers feeding the two BOSS spectrographs \citep{smee13} installed
at the Cassegrain focus of the Sloan Digital Sky Survey 2.5m telescope at Apache Point Observatory \citep{gunn06}.  Each of the six removable cartridges interfaces with a plugplate
system, and contains a full complement of optical fibers bundled into hexagonal IFUs ranging in size from 19 to 127 fibers
along with a set of 12 seven-fiber minibundles for photometric calibration \citep{yan16a}
and 92 individual fibers used for sky subtraction.  In each cartridge the fibers are permanently mounted in a series of v-groove blocks attached to two pseudo-slits that align with the BOSS
spectrograph slitheads.  Since BOSS is a dual-beam spectrograph a dichroic beamsplitter divides the light into blue ($\lambda\lambda 3600-6300$) and red ($\lambda\lambda 5900-10,300$) cameras.
In order to mitigate the effects of atmospheric differential refraction on the effective sampling of the MaNGA fiber bundles
\citep[see][]{law15} each plate is typically observed in sets of three 15-minute dithered exposures
with similar seeing, transparency, and hour angle,
and repeated as necessary 
on multiple MJDs (modified Julian dates)
to reach a target effective depth.

The main MaNGA survey galaxies are drawn from a flat mass distribution in the range $M_{\ast} = 10^9 - 10^{11} M_{\odot}$, with subsamples reaching 1.5 and 2.5 effective radii and a `color-enhanced' subsample designed to obtain
sampling of sparser regions of color-magnitude space
\citep[see discussion by][]{wake17}.
Since its original conception as a dark-time galaxy survey \citep{law15,yan16b}, the MaNGA program has since grown to also encompass a variety of ancillary programs (e.g., observations of M31, IC342, and the Coma cluster) 
as well as a bright-time survey of a large library of stellar spectra \citep[MaStar;][]{yan19}.
The MaNGA Data Reduction Pipeline (DRP) has thus also evolved substantially 
since the initial DR13 public data release (v1.5.4) described by \citet{law16}
to accommodate both this new observing mode and a variety of improvements that have been made
for the main galaxy program.

In brief, the DRP extracts individual fiber spectra from the four detectors via row-by-row optimal extraction, performs sky subtraction and flux calibration  \citep{yan16b} using dedicated calibration fibers and fiber bundles,
and resamples each calibrated spectrum onto a common wavelength grid.  While the majority of these algorithms are identical 
between MaNGA and MaStar, some minor
differences exist since the targets range over 20 magnitudes in brightness (from $g \sim 6$ for bright stellar library targets to  $g \sim 26$ arcsec$^{-2}$ for observations
of intracluster light in the Coma cluster \citep{gu18}) and from dark-sky conditions to mere degrees away from the full moon.
Additionally, for MaNGA galaxy observations the DRP also combines the individual fiber spectra into a composite rectified data cube,
while for MaStar stellar targets the DRP extracts a composite one-dimensional spectrum corrected for atmospheric differential refraction and
geometric fiber-bundle losses from the individual fiber spectra.
As discussed by \citet{law16} the DRP is written almost entirely in IDL, with some C bindings for runtime optimization.

\begin{deluxetable}{lllll}
\tablecolumns{5}
\tablewidth{0pc}
\tabletypesize{\scriptsize}
\tablecaption{MaNGA Data Releases and Pipeline Versions}
\tablehead{
\colhead{Version} &  \colhead{Internal} & \colhead{External} & \colhead{Year} & \colhead{Galaxies}\tablenotemark{a}}
\startdata
3.1.0 & MPL-11 & DR17 & 2021 & 10010\tablenotemark{b} \\
3.0.1 & MPL-10 & ... & 2020 & 9269 \\
2.7.1 & MPL-9 & ... & 2019 & 7823 \\
2.5.3 & MPL-8 & ... & 2018 & 6293 \\
2.4.3 & MPL-7 & DR15 & 2018 & 4532 \\
2.3.1 & MPL-6 & ... & 2017 & 4529 \\
2.1.2 & ... & DR14 & 2017 & 2689 \\
2.0.1 & MPL-5 & ... & 2016 & 2691 \\
1.5.4 & ... & DR13 & 2016 & 1330 \\
1.5.1 & MPL-4 & ... & 2015 & 1329 \\
1.3.3 & MPL-3 & ... & 2015 & 624 \\
1.1.2 & MPL-2 & ... & 2014 & 118 \\
1.0.0 & MPL-1 & ... & 2014 & 58
\enddata
\tablenotetext{a}{Number of unique galaxy targets (discounting special Coma, M31, IC342, and globular cluster targets) with high quality data cubes.}
\tablenotetext{b}{Based on preliminary reductions of the final
survey data}
\label{versions.table}
\end{deluxetable}

In Table \ref{versions.table} we list the versions of the MaNGA pipeline data products available both internally to the 
SDSS collaboration (through MaNGA Product Launches, i.e. `MPL') and
externally to the broader astronomical community (through Data Releases, i.e. `DR').
Some of the changes made for DR13, DR14, and DR15 have already been described by
\citet{dr13}, \citet{dr14}, and \citet{dr15} respectively.
Compared to the DR13 pipeline described by \citet{law16}, the major changes that have been made to the DRP include:

\begin{itemize}
\item Visual yearly inspection of all IFU exposures failing comparisons against established SDSS broadband preimaging data.  Identification
and flagging of IFU frames affected by terrestrial satellite trails allows recovery of high-quality composite data cubes from many previously
flagged as not science-quality.

\item More extensive identification and masking of foreground stars via Galaxy Zoo 3D \citep{masters20}

\item Production of composite single-object spectra for MaStar stellar library targets (v2.0.1), and deredshifting of the resulting spectra (v2.3.1) to the stellar rest frame
\citep[see details given by][]{yan19}

\item Inclusion of full spatial covariance matrices for the galaxy data cubes (v2.0.1).

\item Modification of straylight and bias routines to reduce systematics effects highlighted by ultra-deep observations for the Coma cluster ancillary program \citep[][v2.3.1]{gu18, gu20}

\item Adoption of the BOSZ flux calibration templates \citep{bohlin17} instead of the Kurucz model atmospheres \citep{gray94} \citep[v2.3.1; see][]{yan19}.

\item Adjustments to the IFU fiber bundle metrology to compensate for a $\sim 2.5$\% scale error in laboratory measurements (v2.5.3, v2.7.1; see \S \ref{metrology.sec})

\item Adoption of the \citet{fitz99} extinction curve for standard star calibrations instead of the \citet{od94} curve (v2.5.3).

\item Modification to handle short (5-300 second) exposures for bright MaStar targets (v2.5.3).

\item Addition of special processing to model, subtract, and flag data affected by a bright `blowtorch' artifact resulting
from an electronics failure in the r1 detector during the final year of the survey (v2.7.1, v3.0.1; see \S \ref{blowtorch.sec}).

\item Substantial revisions to the spectral LSF estimation affecting the recovered galaxy velocity dispersions (see discussion in \S \ref{dr15.sec}).

\end{itemize}

The most recent version of the MaNGA DRP data products (MPL-10) consists of all MaNGA and MaSTAR plates completed up to MJD 58933 (March 25 2020).  As determined from the MPL-10
drpall summary file (drpall-v3\_0\_1.fits) it contains 10,529 data cubes corresponding to 630 plates, and 33,360 MaSTAR spectra (26618 unique stars) across
1534 plates.  Of the 10,529 data cubes, 9556 represent galaxies (i.e., discounting the Coma, IC342, M31, and globular cluster ancillary programs
and a few other non-galaxy special plates).  
A small number of these data cubes (162) are flagged by the pipeline as `DONOTUSE' for science based on significant differences between the MaNGA photometry and prior SDSS imaging.  These differences can be due to, e.g., 
bad focus (in which an IFU partially fell out of the plate during observations), unmasked cosmic rays, satellite trails, supernovae, etc.  Of the remaining 9394 galaxy data cubes there are 9269 unique galaxies, roughly 100 of which have two or more independent observations.

The MaNGA MPL-10 data products are available internally to collaboration members both in flat FITS file form\footnote{http://www.sdss.org} and via a python-based API and web application\footnote{https://sas.sdss.org/marvin/}
\citep[Marvin;][]{cherinka19}.  Similarly, DR15 data products are available publicly via the same sources.


\section{Spectral LSF in the MaNGA DRP}
\label{drp.sec}

In order to measure the astrophysical stellar or gas velocity dispersion, e.g. $\sigma_{\Ha}$ from the 
observed \Ha\ emission line profile in a given spaxel, it is critical to have
accurate and precise knowledge of the spectral line spread function
(i.e., the projection into the wavelength domain of the detector point spread function convolved with the tophat pixel sampling).
Since MaNGA uses spectrographs mounted at the Cassegrain focus of the SDSS 2.5m telescope, the LSF 
modulates due to time-variable gravitational flexure in the fibers, camera optics, and detector focal plane; these temporal variations complicate efforts to obtain a robust LSF measurement.

In this section, we discuss the MPL-10 approach taken by the DRP to measure the initial LSF
from calibration arc lamp spectra (\S \ref{arc.sec}), account for pixel
broadening (\S \ref{prepost.sec}), adjust the measurements to match cotemporal night sky features (\S \ref{sky.sec}),
and account for spectral resampling (\S \ref{rectif.sec}) and IFU cube building algorithms (\S \ref{cubes.sec}).
Overall differences from
the initial MaNGA data products provided in previous data releases are summarized in \S \ref{dr15.sec}.


\subsection{Arc-line Model}
\label{arc.sec}

The spectral LSF of the MaNGA data is first estimated using
4-second observations of a Neon-Mercury-Cadmium arc lamp spectrum taken at the beginning of each series of MaNGA observations of a given plate and roughly every 2 hours thereafter.
This arc frame provides a well-populated series of bright unresolved emission lines spanning the wavelength range of the instrument.
Since the wavelength zeropoint is curved along the pseudo-slit \citep[see, e.g., Fig. 22 of][]{law16} this means that a given arc line lies at a range of different `pixel-phases'
(i.e., centroid locations within a pixel) for different fibers, and 
as the relative fiber-to-fiber wavelength
solution is accurate to better than 0.024
pixels rms \citep[see \S 10.3 of][]{law16}
it is therefore possible 
to combine the observed spectra from multiple fibers to obtain a super-sampled realization of the arc line profile.

\begin{figure}
\begin{center}
\includegraphics[width=0.9\columnwidth]{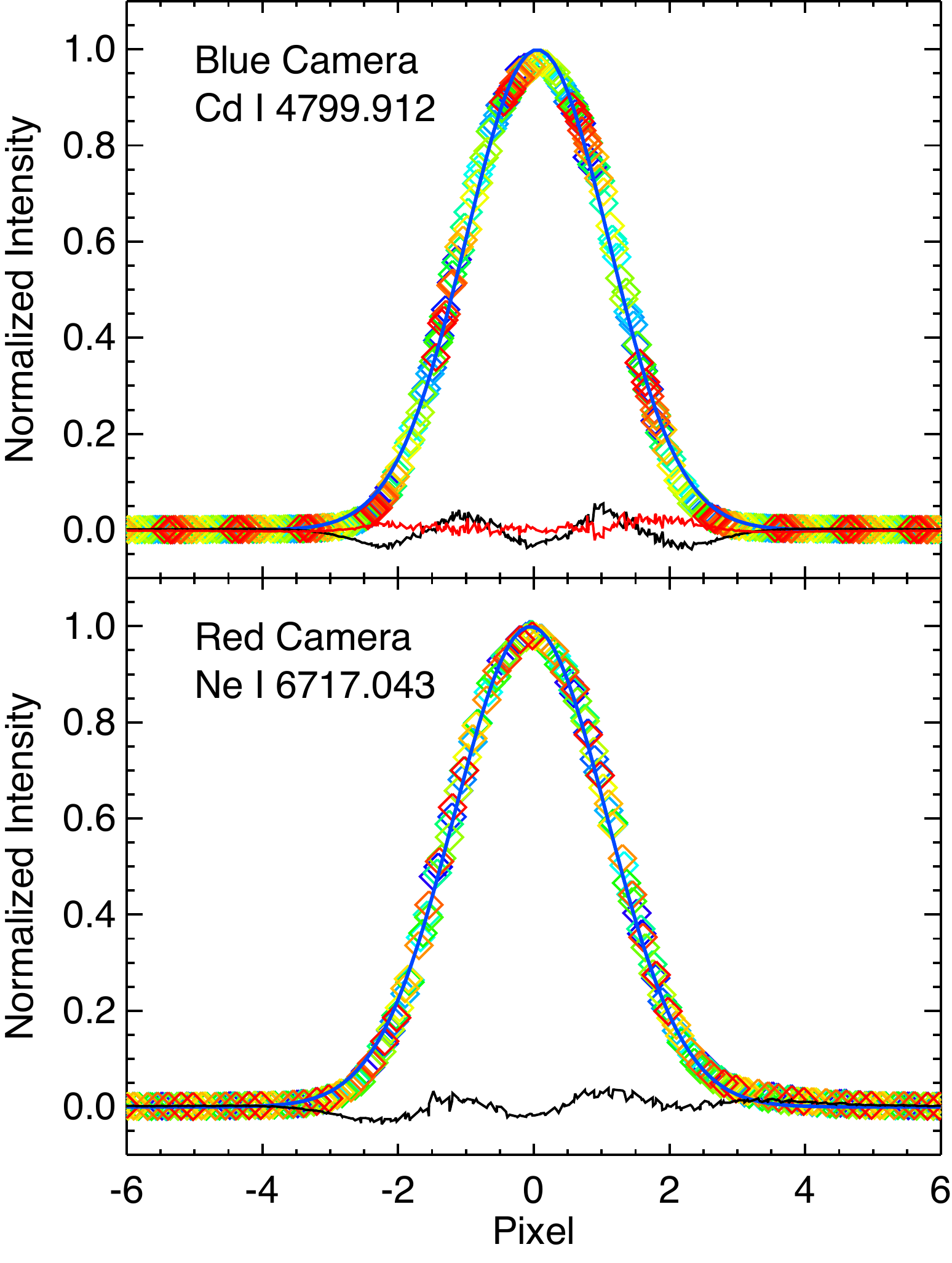}
\end{center}
\caption{Super-sampled observational arc-line profile for characteristic spectral lines in the blue and red cameras.  Colored points represent flat-fielded spectra for 
$\sim 30$ fibers in a single v-groove block for which the LSF is constant across fiberid to $< 1\%$ but the change in zeropoint of the wavelength solution results in the
arc line falling at a variety of different pixel phases for different fibers (color-coded by fiber number).  The solid blue line in both panels represents a Gaussian model fit to the observed data points similar to that used by the MaNGA DRP to describe the line profile, while the solid black line shows the residual difference between the observational data points and the Gaussian model.  These 
residuals demonstrate that the true line profile is modestly platykurtic due to the circular image
of the undispersed optical fiber spot; for comparison a model line profile composed of a gaussian
convolved with a tophat response function can reproduce the observations with negligible residuals (top panel, red solid line).
}
\label{arcline.fig}
\end{figure}

As shown in Figure \ref{arcline.fig}, these line profiles are well described by a simple Gaussian model for both the blue and red cameras across a wide range of wavelengths.
While the residuals from the simple Gaussian fit show evidence for intrinsic kurtosis in the line profile (consistent with expectations for the convolution of a two-dimensional Gaussian with
the circular image of the optical fiber), the peak amplitude of this residual is sufficiently small ($\sim 5\%$ relative to the peak intensity of the line) that it is expected
to have negligible effect on our analysis.\footnote{Although some early commissioning observations
showed optical coma at the longest wavelengths in the red camera of spectrograph 1 this was resolved by realignment of the camera
optics prior to the start of the majority of the survey.}  We therefore fully characterize the shape
of the LSF by a single value $\omega$\footnote{Typically the instrumental LSF is represented as $\sigma_{\rm instr}$, but to avoid a proliferation of subscripts we assign the LSF its own variable.}
giving the $1\sigma$ width of a Gaussian profile fit to the observed pixel values (after first subtracting off the small continuum signal using the median in a 100-pixel window
surrounding the line).

\begin{figure*}
\begin{center}
\includegraphics[width=\textwidth]{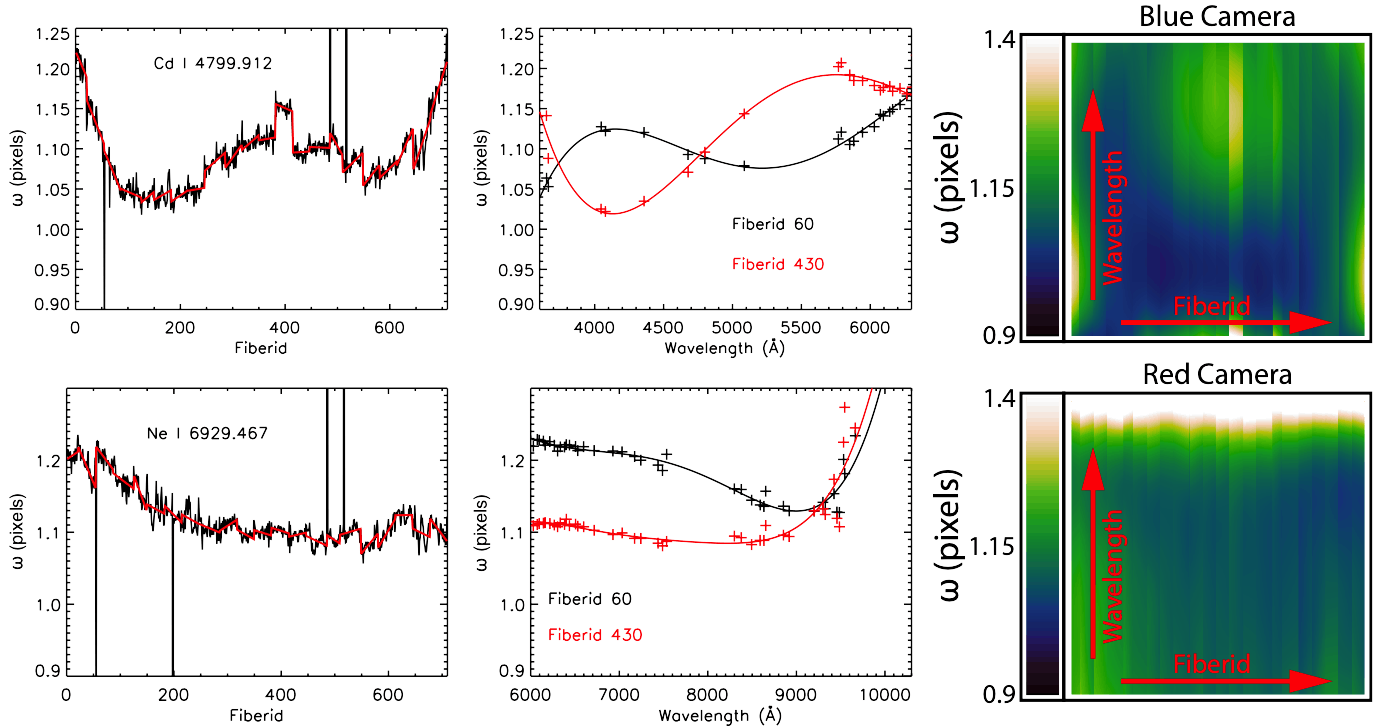}
\end{center}
\caption{Estimates of the MaNGA LSF ($\omega$)
derived from observations of Neon-Mercury-Cadmium calibration lamps.   Left-hand panels: Measured
values of $\omega$ (black line) as a function of fiberid for the indicated lamp line in the blue (top row) and red (bottom row)
cameras.  
Extreme outliers are due to failures in the fitting routine caused by cosmic rays, detector artifacts, and similar effects.
The solid red line shows the fit assuming linear variation within each v-groove block and allowing for arbitrary jumps
between blocks.  Middle panels: Measured values of $\omega$ as a function of wavelength for two example fibers (fiberid 60/430, black/red points
respectively) in each camera.  The solid black and red lines represent the polynomial traceset fit to the observations.
Right-hand panels: Polynomial traceset fits to $\omega$ evaluated across all fiberid  and wavelengths
in the scientifically relevant sections of the detectors (3600-6300 \AA\ and 6000-10,300 \AA\ for the blue/red cameras respectively).
All data shown here are taken from exposure 204255, plate 7960, MJD 57280 (spectrograph 1, cartridge 4). 
}
\label{arcfits.fig}
\end{figure*}

As illustrated by Figure \ref{arcfits.fig} $\omega$ varies over each of the four detectors as a complicated function of both wavelength and fiberid along the 
pseudo-slit.  Similarly, it can change from exposure to exposure with varying telescope/spectrograph focus, gravitational flexure, and changing
observing conditions.  The DRP therefore constructs a model of $\omega$ in pixel units using the individual arc lines in each calibration frame that will be used
as the base calibration for nearby science exposures (the median science exposure is within 34 minutes of the nearest arc frame, 86\% of exposures are within 1 hour, and 99.5\% are within 2 hours).

First, we assume that any
variation in $\omega$ for a given arc line should be approximately linear within a given v-groove block of fibers mounted to the pseudo-slit.
This is because all fibers in a given block will have a common telecentricity with common alignment errors, and should vary in profile only
smoothly with the gradual curvature of the slithead.  Figure \ref{arcfits.fig} (lef-hand panels) shows that this is the case; while individual
measurements for a given fiber are noisy, they describe smooth well-defined trends within a given block with discrete jumps
between adjacent blocks corresponding to alignment differences in their mounting on the pseudo-slit.  
We therefore replace the individual measurements of each arcline in a given fiber with the linear polynomial fit to the fibers in each block; 
this polynomial fit reduces the typical $\omega$ uncertainty by a
factor of about $\sqrt{30}$ (as there are roughly 30 fibers in each block), corresponding
to an improvement
from $\sim$ 1\% to $\sim 0.2$\%
in the wavelength range $\lambda \lambda 6500-7000$ \AA.
This replacement also
has the added benefit of allowing us to be robust against occasional critical failures of the Gaussian-fitting algorithm (resulting, e.g., from bad pixels
or cosmic rays).

Next, we assume that $\omega$ within each fiber should vary smoothly as a function of wavelength within the range $\lambda \lambda 3500-6300$ \AA\ (blue cameras)
and $\lambda \lambda 5900 - 10300$ \AA\ (red cameras).  We therefore fit the linearly-interpolated $\omega$ in each fiber
with an $n$th-order polynomial, where $n=5/6$ for the blue/red cameras respectively.\footnote{These orders are determined empirically to
be the minimum necessary to fit the observed variation.}
As illustrated in Figure \ref{arcfits.fig}  (middle panels), these trace-sets can be evaluated throughout the entire MaNGA wavelength range, and do a good
job of reproducing the observed widths at individual arc lines.  In the $6000-7000$ \AA\ range the density of arc lines is particularly high, and  $\omega$ particularly 
slowly varying with wavelength; the scatter of individual arc lines about the model relation suggest that the overall uncertainty of the fit in the \Ha\ wavelength
regime is about 0.5\%.

Figure \ref{arcfits.fig} (right panels) illustrates the resulting arc-lamp model for $\omega$ across the blue and red cameras; we note
that while $\omega$ in the red cameras is relatively flat as a function of both fiberid and wavelength (rms $\sim$ 0.01 pixels below 9000 \AA) $\omega$ in the
blue cameras shows significantly more structure
(rms $\sim$ 0.03-0.04 pixels between 4000-6000 \AA), corresponding both to the overall curvature of the focal plane and global alignment differences between blocks of fibers. 
The details of this structure are cartridge-dependent since the slithead on each of the six cartridges
has its own mechanical alignment.
Generally, however, fibers $\sim$ 25\% and 75\% of the way along each slit have 
up to a factor $\sim 2$ lower and more constant $\omega$ with wavelength, while fibers near the middle
or ends of the slit show larger variations.

\subsection{Pixel Broadening}
\label{prepost.sec}

As indicated by Figure \ref{arcfits.fig},
MaNGA is nearly critically sampled since spectrally unresolved arc lines typically 
have a $1\sigma$ width of about $\omega = 1.0-1.2$ pixels (2.4-2.8 pixels per FWHM).
As discussed extensively by \citet[][see their Figure 16]{robertson17} such pixels are sufficiently large compared to the 
intrinsic line profile delivered by the telescope/spectrograph optics that the convolution of the intrinsic LSF with the top-hat response of the detector pixels
broadens the overall line profile and must be taken into account.
Our post-pixellized measurements of $\omega$ that were obtained by simply evaluating a Gaussian profile model at the midpoint of each pixel
are therefore systematically biased relative to the `true' intrinsic instrumental dispersion
before convolution with the pixel response function (i.e., the pre-pixellized $\omega$).

\begin{figure}
\begin{center}
\includegraphics[width=\columnwidth]{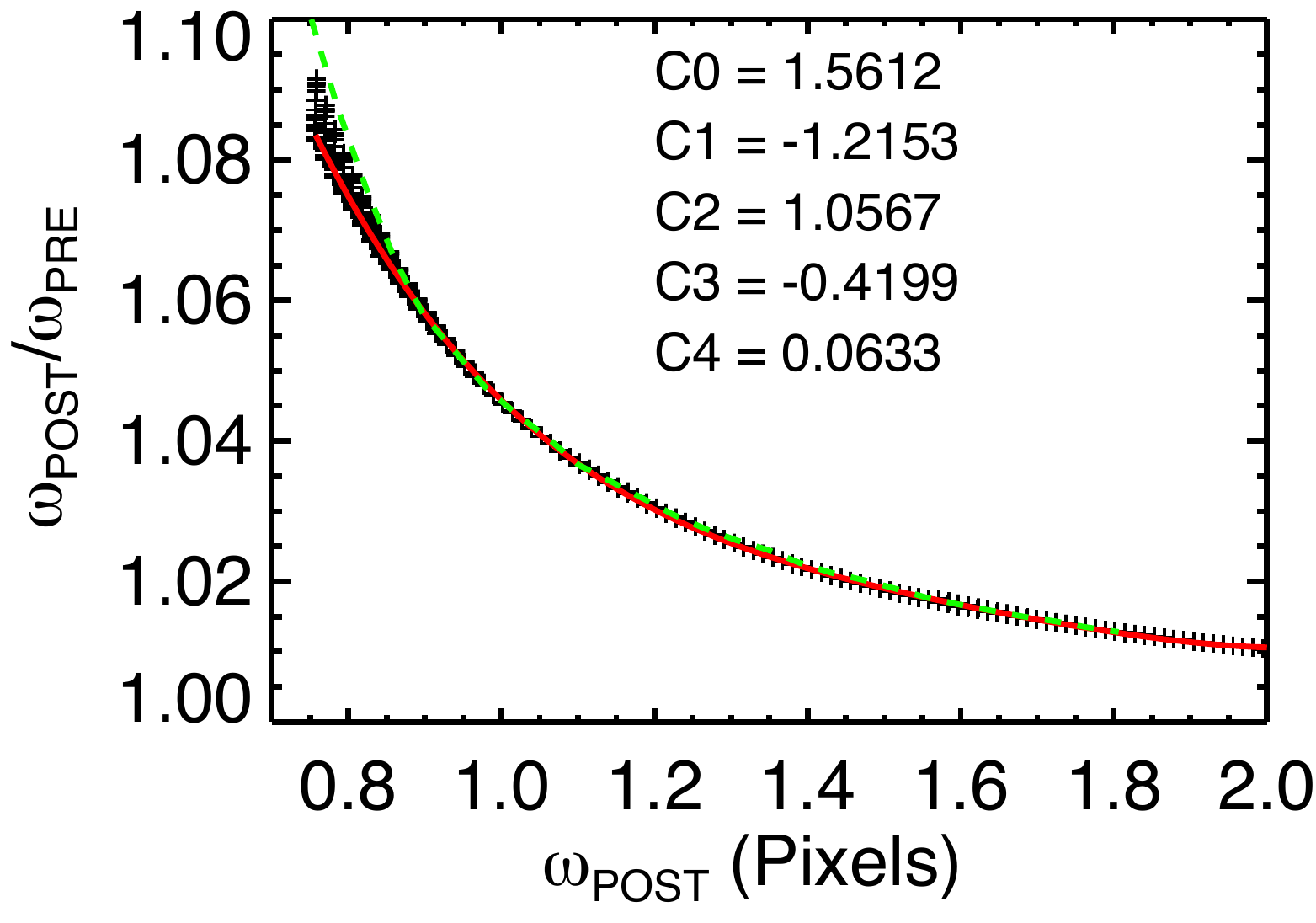}
\end{center}
\caption{Systematic overestimate of the true spectral LSF ($\omega_{\rm PRE}$) as a function of the LSF derived from fitting a Gaussian profile ($\omega_{\rm POST}$) to 
a series of Monte-Carlo generated mock arc-lamp spectra.  Each black '+' symbol represents a single test with different pixel phase; differences between pixel phases are only
apparent at the smallest $\omega_{\rm POST} \sim 0.8$ pixels, at which the peak-to-peak scatter is about 1\%.  In the typical MaNGA range ($\omega_{\rm POST} =$ 1.02/1.11/1.31 pixels for the 5th/50th/95th percentile of the MPL-10 distribution)
pixel phase effects are less than 0.1\%.   The solid red line represents the 4th-order polynomial fit to this relation used by the DRP; the coefficients of this fit are given in the inset text.
Also shown for comparison (dashed green line) is the similar relation derived by \citet{robertson17}.}
\label{prepost.fig}
\end{figure}

We compute the magnitude of this effect using Monte Carlo simulations in which we constructed 10x oversampled Gaussian models of known width, convolved them with
the pixel response function at a variety of pixel phases, and then measured the resulting profiles using the commonly-available Gaussian-fitting techniques that simply
evaluate the Gaussian function at the pixel midpoints.
As illustrated in Figure \ref{prepost.fig}, post-pixellized widths are systematically broader than the pre-pixellized values by $\sim 1-10$ \% and define an extremely tight
mathematical relation in which the pixel sampling phase drives the scatter but is of negligible importance ($<0.1$ \% ) in the range of line widths observed by MaNGA.
The DRP therefore computes pre-pixellized estimates of the MANGA LSF ($\omega_{\rm PRE}$) from the measured post-pixellized values ($\omega_{\rm POST}$) using
a 4th-order polynomial fit to this relation (red line in Figure \ref{prepost.fig}).

As motivated and discussed in \S \ref{dap.sec}, the MaNGA Data Analysis Pipeline (DAP) adopts these pre-pixellized estimates of $\omega$ and rigorously account for pixel convolution. 
However, since many third-party analysis routines ignore pixel convolution and instead rely on simple Gaussian-fitting approximation the DRP provides both $\omega_{\rm POST}$
and $\omega_{\rm PRE}$ with all of the survey data products.

\subsection{Sky-line Model}
\label{sky.sec}

In practice, differences in the telescope focus (due to, e.g., changing weather conditions), gravitational flexure of the spectrographs, and various other effects
mean that the LSF derived from arc-lamp exposures are only an approximation to the actual LSF of any given science exposure.  We therefore use the well-known bright sky
emission line features to refine the LSF estimate for each individual science exposure.\footnote{Such refinements are not possible for the shortest MaStar exposures in which even strong
skylines are relatively faint, but since these exposures are obtained much closer in time to the calibration exposures focus drifts are much less common than in longer MaNGA observations
that can differ by an hour from the calibration exposures.}

Unlike the arc spectra, however, the night-sky emission features are not ideally distributed in wavelength (there are very few bright skylines at blue wavelengths)
and they can frequently be biased by continuum emission and blends of multiple atomic and/or molecular transitions.
While such blending may be weak, even weak blending can bias apparent measurements of the LSF at the few-percent level.
Rather than rederiving the LSF solution from the skylines, the DRP therefore uses them to simply make low-order adjustments to the arcline model.

\begin{deluxetable}{ll}
\tablecolumns{2}
\tablewidth{0pc}
\tabletypesize{\scriptsize}
\tablecaption{Night Sky Calibration Lines}
\tablehead{
\colhead{Wavelength (\AA)} &  \colhead{Transition}}
\startdata
  4046.56\tablenotemark{a}    & Hg I\\
  4358.33\tablenotemark{a}    & Hg I\\
  5460.94\tablenotemark{a}    & Hg I \\
  5577.339   & O I\\
  6300.304\tablenotemark{a}   & O I \\
  6363.776    & O I \\
  7571.75    & OH  \\
  7794.12    & OH \\
  7821.51    & OH \\
  8399.16    & OH\\
  8430.17    & OH \\
  8885.83    & OH \\
  8919.61    & OH \\
  9439.65    & OH \\
  9872.13    & OH\\
  10124.01  & OH 
\enddata
\label{skylines.table}
\tablenotetext{a}{Used only for adjustment of the wavelength solution, not adjustment of the LSF.}
\end{deluxetable}

\begin{figure}
\begin{center}
\includegraphics[width=0.8\columnwidth]{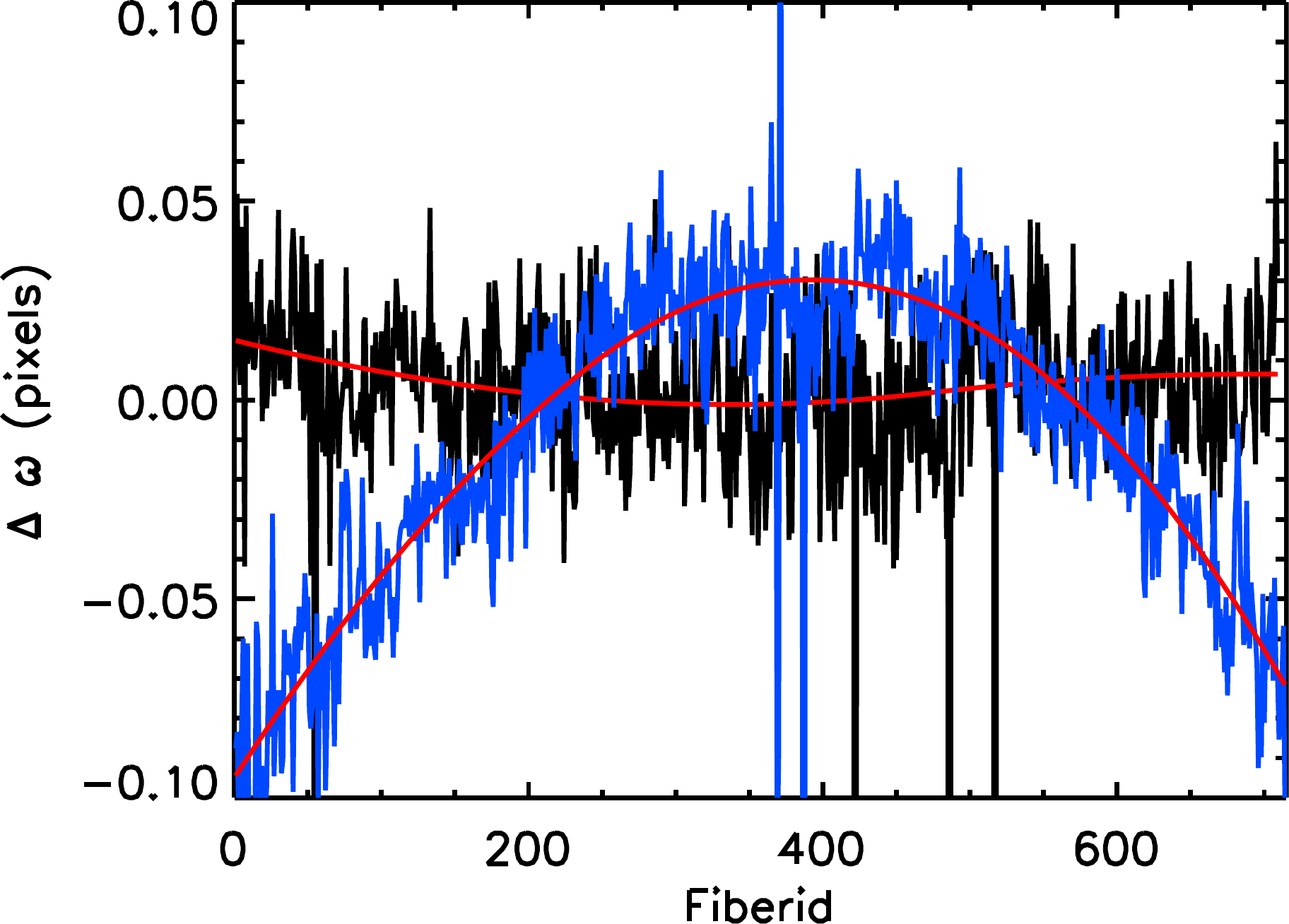}
\end{center}
\caption{Difference $\Delta \omega$ between the measured \oone\ $\lambda 5577.339$ skyline LSF and the arc-lamp model evaluated at the skyline wavelength for a typical
case (7960-57280-204255-b1; black solid line) and for a large-shift case (7960-57458-216817-b2); blue solid line). In each case the solid red line shows the cubic spline fit to the observed values used by the MaNGA DRP. }
\label{skydiff.fig}
\end{figure}

Using the extracted, flat-fielded science frame spectra we fit each skyline in our list of reliable lines (see Table \ref{skylines.table}) with a post-pixellized Gaussian model that includes a linear
polynomial term to account for wavelength gradients in the sky continuum level.  We then compute the difference
$\Delta \omega = \omega_{\rm sky} - \omega_{\rm arc}$, where $\omega_{\rm arc}$ is the arc-line LSF model evaluated at the wavelength of the skyline features.  Even in the cases for which the sky and arcline measurements differ substantially, the difference $\Delta \omega$ between the measurements shows an extremely smooth variation
along the slit with no noticeable block-to-block jumps and only statistical noise from the individual fiber measurements (Figure \ref{skydiff.fig}).  We therefore fit (unmasked) values of 
$\Delta \omega$ with a cubic basis spline with breakpoints every 150 fibers to obtain our initial estimate of the difference between skyline and arcline LSF models.

\begin{figure*}
\begin{center}
\includegraphics[width=\textwidth]{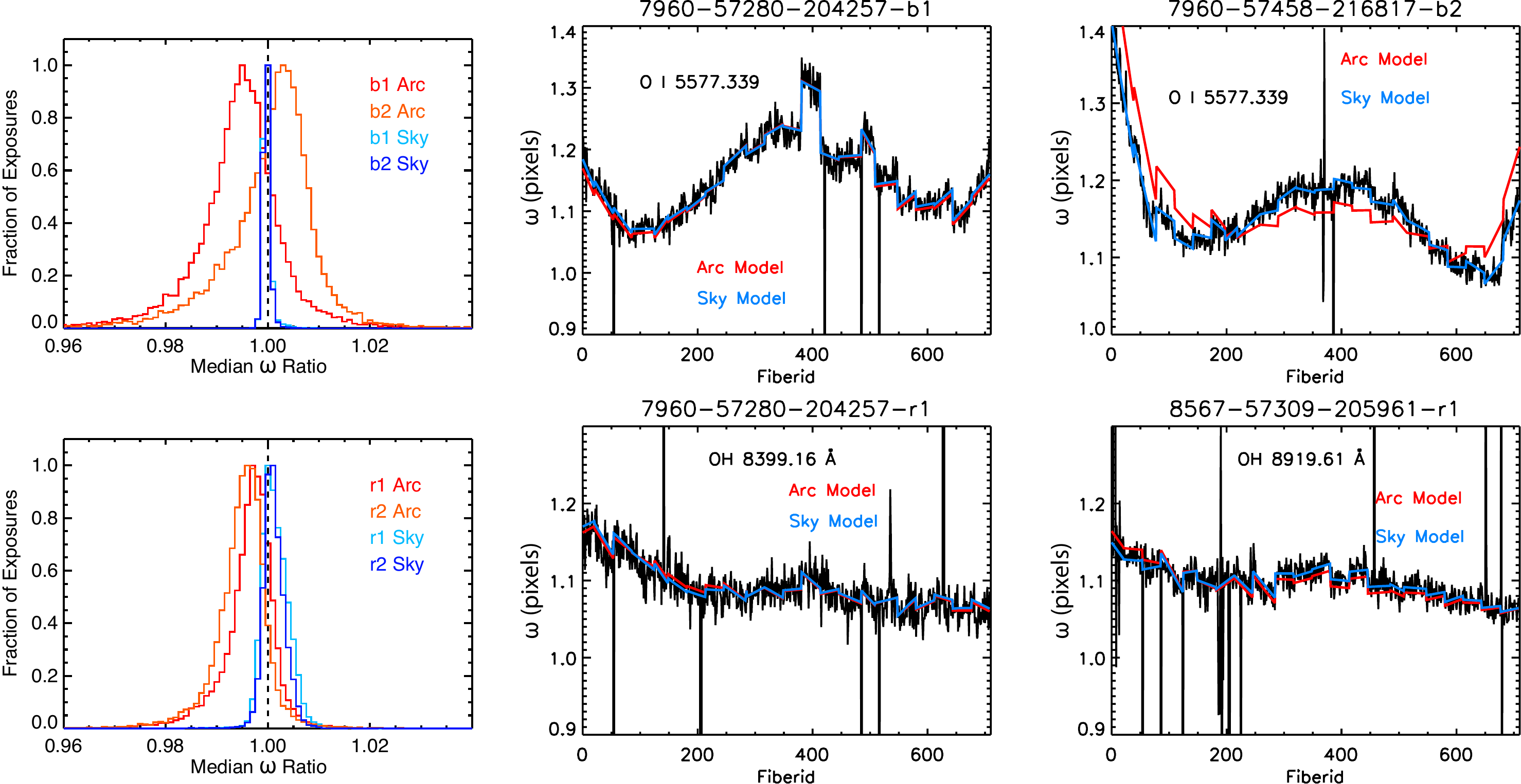}
\end{center}
\caption{Left column: Histogram of 
the median ratio per exposure between the arcline and skyline LSF models, and between the skyline
LSF models and the individual skyline measurements for the $\sim 22,000$ individual science
exposures in MPL-10. 
While the width of the arcline histograms is dominated by the differences between the arc model and the skyline measurements, the skyline histogram width
is dominated by the uncertainty in the skyline measurements.  Note that the blue histograms are narrow than the red as they are based on a single skyline.  Middle column: Measured
skyline widths (black solid line) as a function of fiberid compared to the arcline and skyline models (solid red and blue lines respectively) for a typical case in the blue and red cameras.
Right column: As middle column, but for a large-shift example.  Note that the skyline model does not always go through the middle of the observed skyline measurements for a given line
in the red camera because individual lines can have systematic biases due to blending and continuum-fitting problems.}
\label{skylines.fig}
\end{figure*}

In the blue camera ($\lambda\lambda = 3600-6300$ \AA) only the bright \oone\ $\lambda 5577.339$ skyline is deemed to be a reliable LSF calibrator since all of the other lines used to adjust the MaNGA wavelength solution are marginally blended at the few-percent level, and the $\Delta \omega$ derived from this line is therefore assumed to be constant for all wavelengths.
In the red camera ($\lambda\lambda = 5900-10,300$ \AA) there are multiple skylines, but since we observe no unambiguous trends in $\Delta \omega$ with wavelength we 
simply median combine each of the estimates to obtain our final correction value.
The resulting sky-adjusted LSF models are typically different from the original arc-line models by less than 0.01 pixels, although in some extreme cases can differ by around 0.05 pixels
(Figure \ref{skylines.fig}, middle and right-hand panels respectively).
Based on the $> 22,000$ individual science exposures in MPL-10, we note that the 
arcline model tends to systematically underestimate the observed skyline width by $\sim 0.5$\% on average in the b1, r1, and r2 cameras, and overestimates the skyline width
by about the same amount in the b2 camera (Figure \ref{skylines.fig}, left-hand panels), possibly due
to systematic differences in optical alignment between the cameras.
In contrast, the skyline-adjusted models have negligible systematic offset from the skyline measurements and a significantly smaller width to the distribution that is dominated by the uncertainty
in individual lines.

\subsection{Wavelength Rectification and Dichroic Recombination}
\label{rectif.sec}

As discussed in detail by \citet{law16}, the MANGA DRP processes each camera of data independently up to the point of producing flux-calibrated, sky-subtracted spectra for each fiber.  Once all four
cameras have been thus processed, the DRP stitches together the natively-sampled spectra from the blue and red cameras across the dichroic break to produce final calibrated spectra for each fiber
that cover the entire MaNGA wavelength range.  This is achieved via high-order cubic  basis spline modeling of the blue and red spectra with a tapered inverse variance weighting\footnote{Smoothed
to mitigate the well-known systematic biases that inverse variance weighting can introduce in the median counts of combined spectra.} in the $5900 - 6300$ \AA\ dichroic window to provide a 
smooth transition between the cameras.  This spline model is evaluated on two different output grids; a linear solution with a constant $\Delta \lambda = 1$ \AA\, and a logarithmic wavelength
solution with a constant $\Delta {\rm log} (\lambda/{\rm \AA}) = 10^{-4}$.  While the
linear wavelength solution products are used for some 
MaNGA Value Added Catalogs \citep[e.g., Pipe3D;][]{sanchez16}, the MaNGA
Data Analysis Pipeline (DAP; \S \ref{dap.sec}) uses the products with a logarithmic
wavelength solution.

\begin{figure*}
\begin{center}
\includegraphics[width=\textwidth]{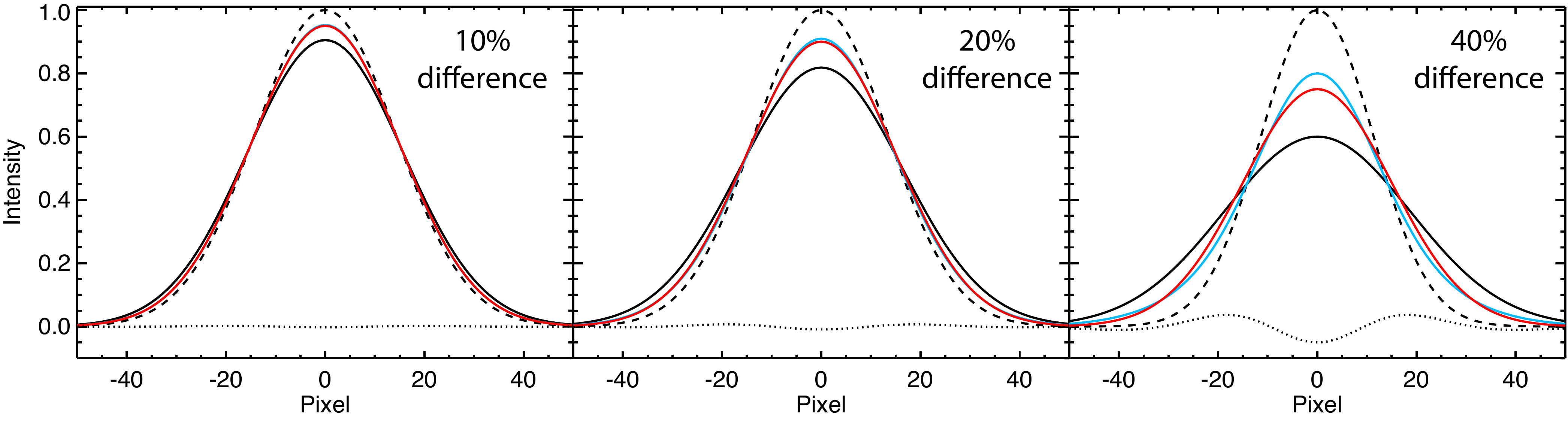}
\end{center}
\caption{Comparison between Gaussian LSF models produced by summing Gaussians of various widths. In each panel the solid and dashed black lines respectively show Gaussian profiles whose widths differ by the indicated amount at fixed integrated line intensity. The blue line indicates the profile produced by summing the solid/dashed black profiles, while the red line indicates the profile of a single Gaussian whose width is given by the average of the first two. The dotted black line shows the residual difference between the true summed profile (blue line) and the approximate gaussian model profile (red line). These residuals are small compared to the known non-gaussianity of the instrumental profile (Figure \ref{arcline.fig}) for the typical
difference in LSF widths ($\lesssim 15$\%) combined by the MaNGA DRP.
}
\label{newfig.fig}
\end{figure*}

The combination of the per-camera LSF vectors onto the final 
rectified wavelength grid uses the same algorithm as for the spectra themselves.
In the $5900 - 6300$ \AA\ dichroic overlap region the
gradual tapering of the weights applied to the blue/red cameras serves to
produce a smooth transition between the LSF solutions of the individual cameras, but
the spectra are nonetheless a composite of individual spectra whose LSF widths
differ from each other by about 15\%.  As we show in Figure \ref{newfig.fig} however,
the non-gaussianity introduced as a result is insignificant, especially compared to the
known intrinsic line profile (Figure \ref{arcline.fig}).

In addition to simply mapping the LSF onto the output wavelength grid however, the
wavelength rectification also
broadens the effective LSF slightly compared to the original spectra that were sampled by the native detector pixels.
We compute the magnitude of this broadening using a series of Monte Carlo simulations for a statistically large grid of Gaussian `comb' spectra, in which artificial spectra with emission lines of known
width are created every 100 \AA\ throughout the MaNGA wavelength range.  These artificial spectra are produced using the wavelength solution of a typical natively-sampled exposure,
combined together across the dichroic to produce composite spectra using the spline algorithm described above, and the widths of the resulting lines then computed via gauss-fitting techniques
to compare to the known input widths.  By repeating this experiment across the 700+ fibers per MaNGA spectrograph, shifting the input line centroids by sub-pixel values to consider ten different
input pixel phases, and covering a range of twelve different input widths from 0.9 to 2.0 pixels we are therefore working with a grid of $> 6$ million simulated lines.

\begin{figure}
\begin{center}
\includegraphics[width=\columnwidth]{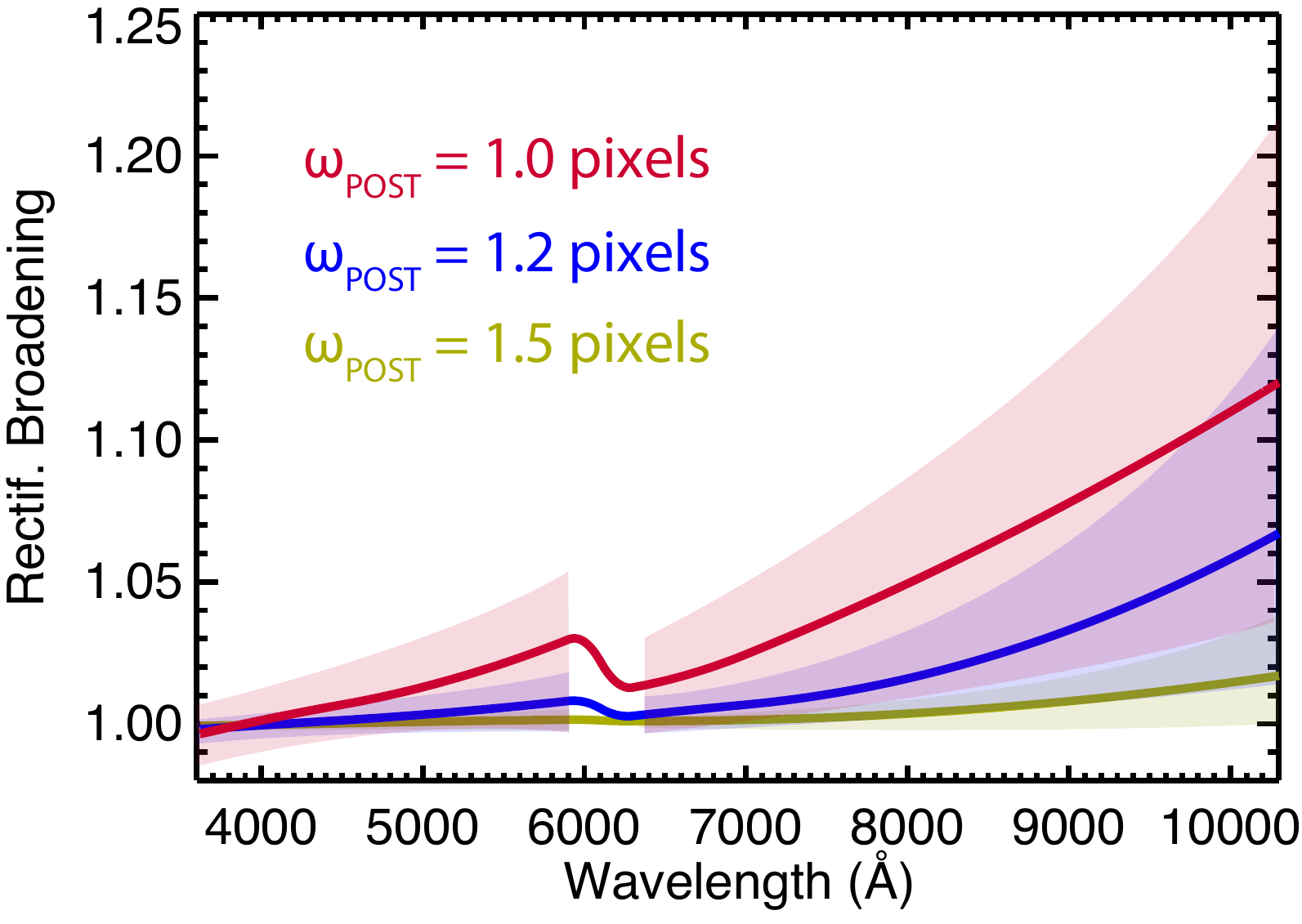}
\end{center}
\caption{Effective LSF broadening factor (i.e., the ratio of the post-pixellized LSF in units of \AA\ measured before and after rectification) introduced by the spectral rectification of the native-pixel blue and red camera spectra onto a fixed logarithmic wavelength grid.  Solid red, blue, and gold lines
show the median broadening as a function of wavelength computed via Monte Carlo analysis of artificial spectra for input widths of $\omega = 1.0, 1.2, and 1.5$ pixels respectively.  In each case the shaded region shows
the $1\sigma$ variation about the mean traced by input lines with different effective pixel phases.  The corresponding curves for the linear wavelength grid (not shown) are generally similar 
except in the 9000 - 10,000 \AA\ range where the logarithmic output grid undersamples the typical input LSF.
}
\label{broadfac.fig}
\end{figure}

As shown in Figure \ref{broadfac.fig}, the effective broadening factor (post-pixel to post-pixel) is a strong function of wavelength, increasing from near unity at the shortest wavelengths in both the blue
and red cameras to a $\sim$ 10\% effect at the longest wavelengths in the red camera for input $\omega_{\rm POST} = 1.0$ pixels.  The exact correction factor is strongly dependent
on the input pixel phase, particularly for values of $\omega_{\rm POST} \leq 1.0$ pixel.  Since any DRP correction to the LSF cannot take pixel phase into account (since this will be different for every
emission line in the science data in a manner that cannot easily be modeled), we fit the median relation as a function of wavelength\footnote{We do not include the dichroic overlap region in our fit since
different LSF widths from the blue and red cameras are being combined here into a single line profile, but instead simply ensure that the spline model smoothly joins between the blue and red
camera solutions.} for each of our twelve input widths with a cubic basis spline
to compute a grid of correction factors for both the logarithmic and linear wavelength solutions.  
The DRP then corrects the composite LSF vectors for each fiber by the appropriate
value interpolated from this reference grid
(applying the same factor to both $\omega_{\rm PRE}$ and $\omega_{\rm POST}$ estimates).
Given the typical range of $\omega_{\rm POST} = 1.1 - 1.2$ pixels in the vicinity of \Ha\, the range of correction factors across different pixel phases indicated by Figure \ref{broadfac.fig}
suggests that the rms uncertainty of the applied correction is typically around 1\%.



\begin{figure}
\begin{center}
\includegraphics[width=\columnwidth]{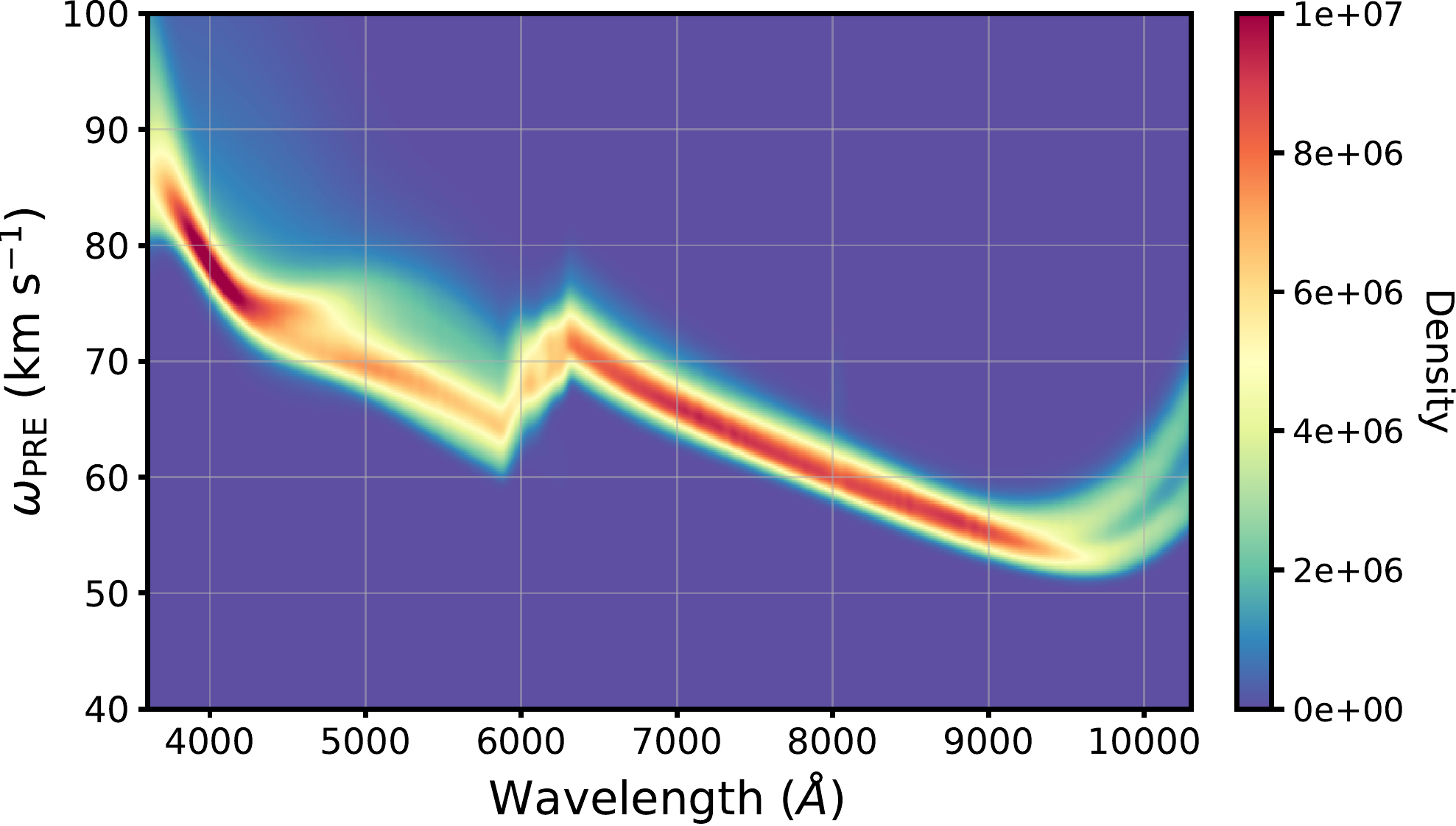}
\end{center}
\caption{Density plot of the pre-pixellized instrumental LSF (in velocity units) reported by the MaNGA DRP for all $\sim 31$ million individual fiber spectra in MPL-10
(for the LOG wavelength solution).
}
\label{rsslsf.fig}
\end{figure}

The corrected pre-pixel instrumental LSF for all individual calibrated exposures ($\sim$ 30 million individual fiber spectra) in MPL-10  are shown in Figure \ref{rsslsf.fig}.  We conclude that the typical instrumental resolution improves from $80-90$ \kms\ at the bluest wavelengths to about 55 \kms\ at the red end
of the MaNGA wavelength range, with the largest RMS variation between fibers around 5000-6000 \AA\ and the smallest RMS variation of just a few percent around \Ha.

\subsection{Data Cube Construction}
\label{cubes.sec}

The DRP additionally reformats the
calibrated fiber spectra into a rectilinear datacube in which the individual fiber spectra have been coadded to produce a single three-dimensional datacube with two spatial axes and one spectral axis.  Working with these datacubes is significantly easier than working directly with the individual calibrated fiber spectra (provided by the DRP as ``row-stacked spectra'', or RSS) as the latter suffer
from chromatic differential refraction while the spectra in the rectified data cubes are directly
associated with a specific location on the sky.
For this reason, the vast majority of the MaNGA science team has thus used the composite data cubes for science analyses, as does the MaNGA Data Analysis Pipeline (DAP).


However, the algorithm used to construct these datacubes produces complications
of its own.
In addition to introducing strong spatial covariance between adjacent datacube spaxels, as explored in detail by \citet[][Section 9]{law16} and \citet[][Section 6]{westfall19}, the datacube construction also combines spectra with a variety of different spectral resolutions.  Since the effective LSF can vary strongly from fiber-to-fiber, exposure-to-exposure, and night-to-night the range of input resolutions contributing to any given datacube spaxel can be non-negligible.  

\begin{figure}
\begin{center}
\includegraphics[width=\columnwidth]{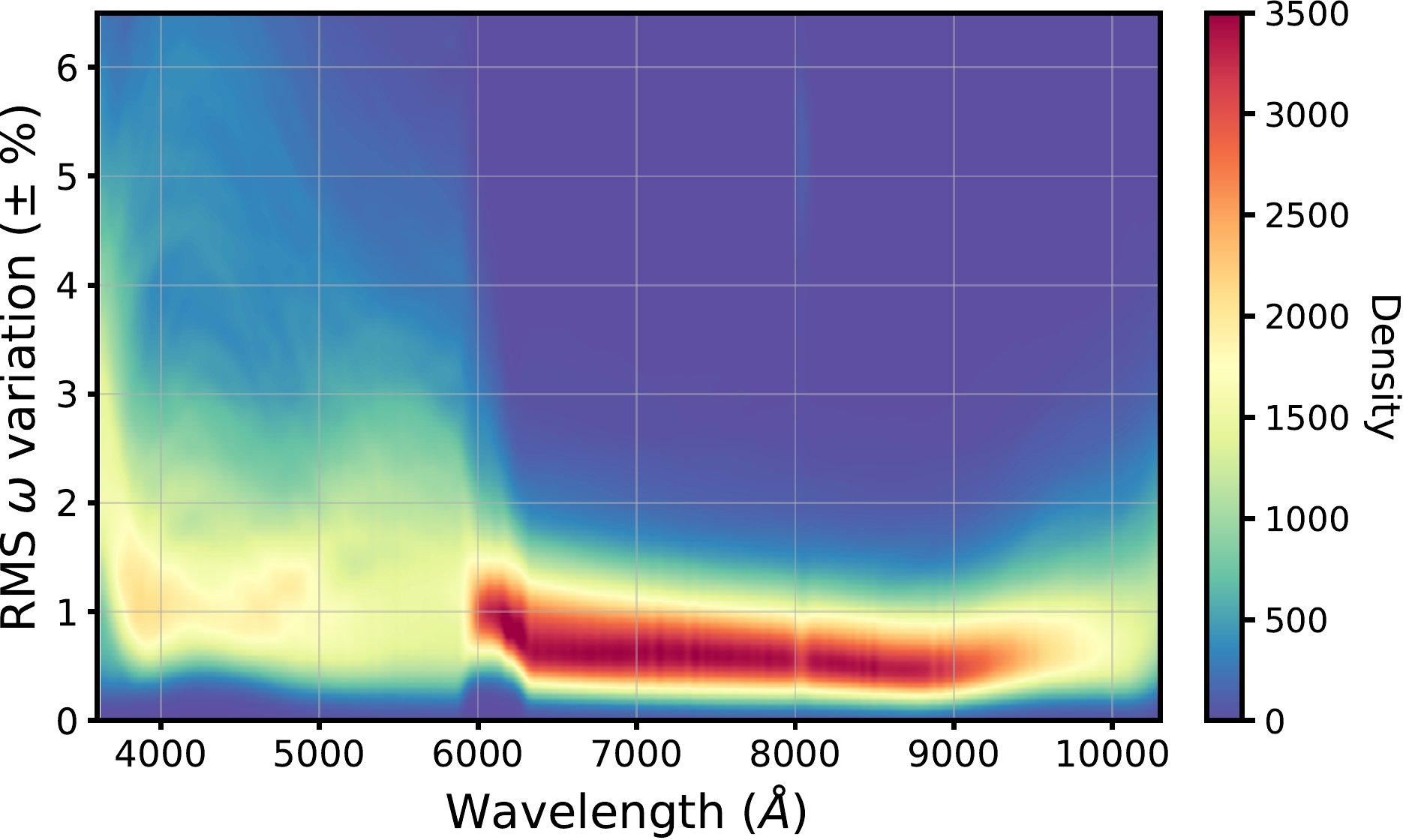}
\end{center}
\caption{RMS variation of $\omega_{\rm PRE}$ as a function of wavelength between all fiber spectra contributing to a given data cube for all 10,523 data cubes in MPL-10.}
\label{lsf_variation.fig}
\end{figure}


Figure \ref{lsf_variation.fig} shows a density plot of the RMS variation in $\omega$ of all fibers contributing to a given datacube as a function of wavelength for all 10,523 data cubes in MPL-10.  As discussed in \citet{law16}, typical IFUs show variability at the 1-2\% level in the blue cameras with rare worst-case outliers at about 10-15\% at some wavelengths for IFUs on the edges of the slit. In contrast, driven by the flatter focal plane across the CCDs the focus in the red cameras is significantly flatter than in the blue, with the majority of all data cubes showing $< 1$\% variability in the component fiber spectra. 
Given this relatively small LSF variability, we avoid the complexity of attempting to convolve all fiber spectra to the same resolution (i.e., the difficulty in making such small adjustments to the {\it wavelength-dependent} resolution, the loss of information by degrading the majority of the datacube spectra, and the introduction of more spectral covariance; \citep[see, e.g.,][]{pace19}) and instead construct an LSF width metric for the combined spectra.
I.e., assuming a Gaussian function for the astrophysical LOSVD, we want a simple
metric that can be used to accurately remove the influence of the LSF on our measurement of the astrophysical velocity dispersion.

To understand the influence of the LSF metric we perform an experiment by
constructing two Gaussian profiles, each with a width defined by perturbing $\omega$ by a small percentage above and below 70 \kms (e.g, we denote a 1\% change as $\delta\omega/\omega = 0.01$).  We then construct an ``observed'' profile by summing these two Gaussian profiles
and convolving the result with a third Gaussian with dispersion $\sigma_{\rm in}$
that represents the LOSVD of the gas in the galaxy.
We then fit a fourth Gaussian to the result, mimicking the typical procedure when fitting galaxy data.  The dispersion of the best-fitting Gaussian is then corrected for the LSF width to produce $\sigma_{\rm out}$; any difference between $\sigma_{\rm in}$ and $\sigma_{\rm out}$ indicates a measurement bias.  The experiment is performed with highly oversampled noiseless profiles so as to explore the {\it intrinsic} bias in each approach.

\begin{figure}
\begin{center}
\includegraphics[width=\columnwidth]{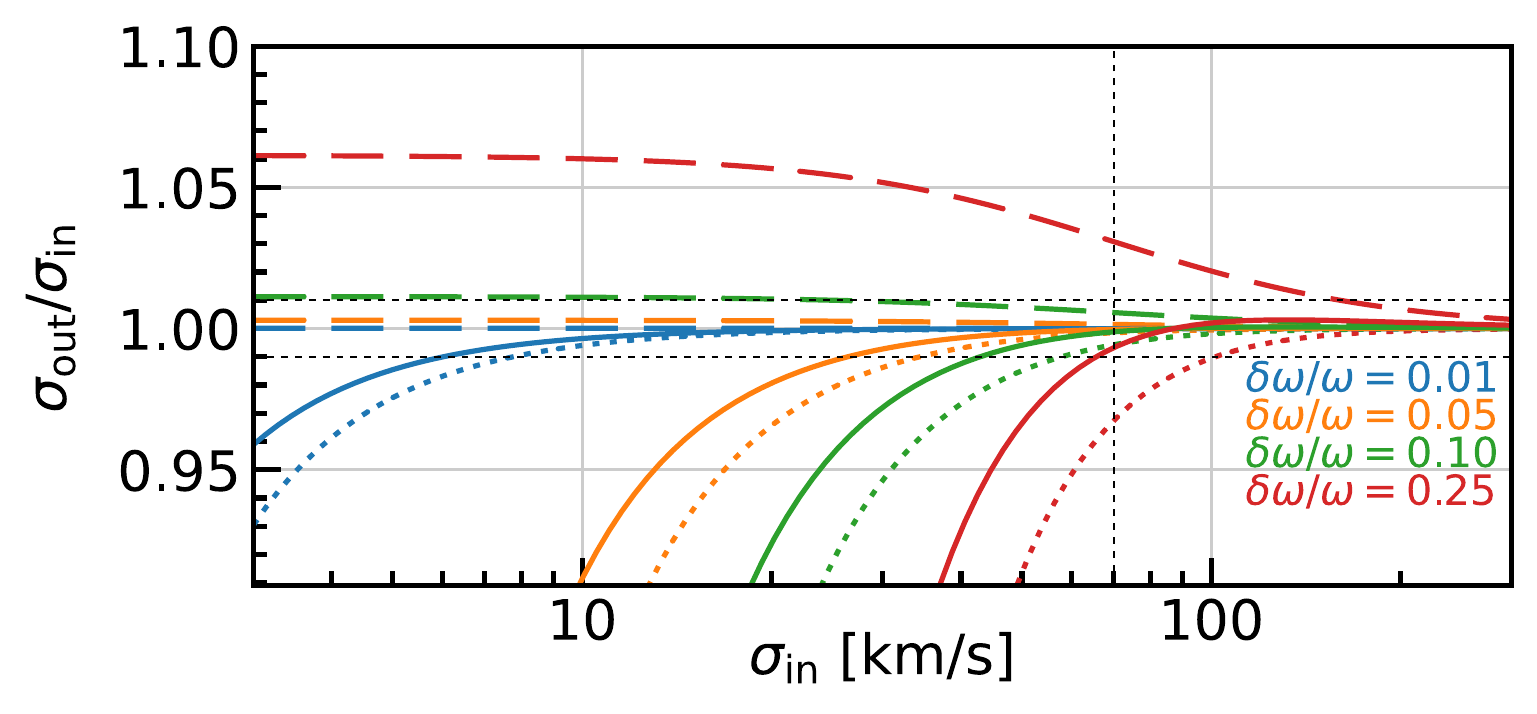}
\end{center}
\caption{Numerical tests of the bias in the recovered astrophysical velocity dispersion ($\sigma_{\rm out}$) compared to a known input value ($\sigma_{\rm in}$) for three different descriptions of a composite LSF width.  Line types indicates the method used to construct the composite LSF width (see text) while line color indicates the variation in the LSFs combined to construct the composite LSF ($\delta\omega/\omega$; see legend).  From Figure \ref{lsf_variation.fig}, we know that the MaNGA LSF variation is typically $\lesssim$1-2\% (blue lines), 
such that the exact definition of the composite LSF width is effectively irrelevant for $\sigma_{\rm in} > 10$ \kms.}
\label{lsfbias.fig}
\end{figure}

In Figure \ref{lsfbias.fig}, we show the results of this test for four values of $\delta\omega/\omega$ (differentiated by line color) and three different 
methods of estimating the combined LSF metric:
(1; dotted lines) The second moment of the summed profile $\omega^2 = (\omega^2_{\rm lo} + \omega^2_{\rm hi})/2$, where $\omega^2_{\rm lo}$ and $\omega^2_{\rm hi}$ are the values for the narrow and broad Gaussian components.  (2; solid lines) The mean $\omega$ of the two components. (3; dashed lines) The width of a new Gaussian profile fit directly to the summed profile.

Figure \ref{lsfbias.fig} demonstrates that when the range in LSF widths for combined spectra is $\lesssim$1\% (true for the majority of MaNGA datacubes, particularly for $\lambda \gtrsim 6000$ \AA) the method used to estimate the combined LSF is largely irrelevant; any bias is $<$1\% for $\sigma_{\rm in} > 10$ \kms\ for all methods.  For larger differences in the LSF widths, fitting the composite LSF to define $\omega$ is generally better than the other two definitions.  However, given that this method requires significantly more computational overhead --- requiring us to construct the composite profile and fit a Gaussian for every pixel with valid spectral data in the full MaNGA dataset --- and the fact that it indeed introduces biases of its own, we instead adopt the simple mean method, which performs modestly better than using the second moment.  Combined with Figure \ref{lsf_variation.fig}, we expect this simple mean method of combining the fiber LSFs will rarely introduce more than a 1 \kms\ bias
for $\sigma \gtrsim 20$ \kms\ even
in the extreme case of 5\% variation in the fiber $\omega$,
and will more typically be $<$0.1 \kms\ (far below the typical uncertainty in the individual measurements; see Figure \ref{uncertainty.fig}). 
As illustrated by Figure \ref{newfig.fig}, differences in the LSF of the input spectra
at such levels will have negligible impact on the overall gaussianity of the composite
spectrum.

We therefore construct LSF datacubes using the same algorithm and weighting scheme as adopted to construct the science data cube.  That is, with the modified Shepard
algorithm the image of the galaxy at a given wavelength slice is a weighted sum of the input fiber spectra at that wavelength, and the LSF `image' is constructed from the same weights
applied to the LSF vectors of the individual fibers.

\begin{figure*}
\begin{center}
\includegraphics[width=\textwidth]{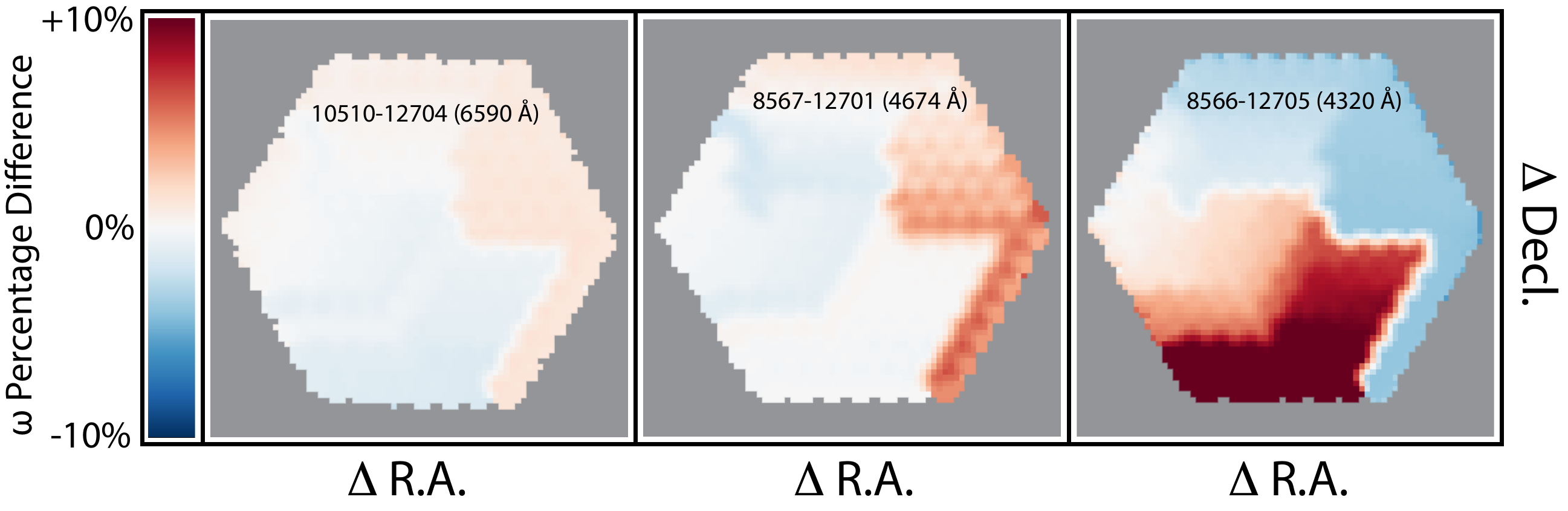}
\end{center}
\caption{Spatial variability in $\omega$ 
for three example data cubes at various wavelengths (left to right: 1\%, 2\%, and 6\%
rms variability).
The visible structure traces the mapping of the IFU fibers to discrete V-groove blocks
on the BOSS spectrograph pseudo-slit \citep[c.f. Figure 9 of][with a horizontal flip in orientation]{drory15}.  Inset text gives the MANGAID and wavelength of each example.}
\label{cubelsf.fig}
\end{figure*}

The MaNGA DRP provides both a summary averaged LSF vector for each data cube (as small variations will be unimportant to science cases studying astrophysical line widths in excess of 100 \kms), as well as a full three-dimensionsal LSF datacube so that downstream analysis programs can use the effective LSF appropriate for each spaxel in the cube.  For illustrative purposes we show in Figure \ref{cubelsf.fig} examples with 1\%, 2\%,
and 6\% LSF variation across the face of the IFU as calculated by the DRP.
Such spatial variation in the effective LSF of integral-field data is well known for slit-type and lenslet-type spectrographs as well \citep[see, e.g., Fig 6 of][]{law18} and must be taken into account when measuring velocity dispersions near the instrumental resolution or below.
As suggested by Figure \ref{lsf_variation.fig}, 99.9\% of  cubes resemble the 1\% or
2\% rms
examples at red wavelengths, while 80\% resemble these examples at blue wavelengths.


\subsection{LSF Differences from Prior Data Releases}
\label{dr15.sec}

\begin{figure*}
\begin{center}
\includegraphics[width=\textwidth]{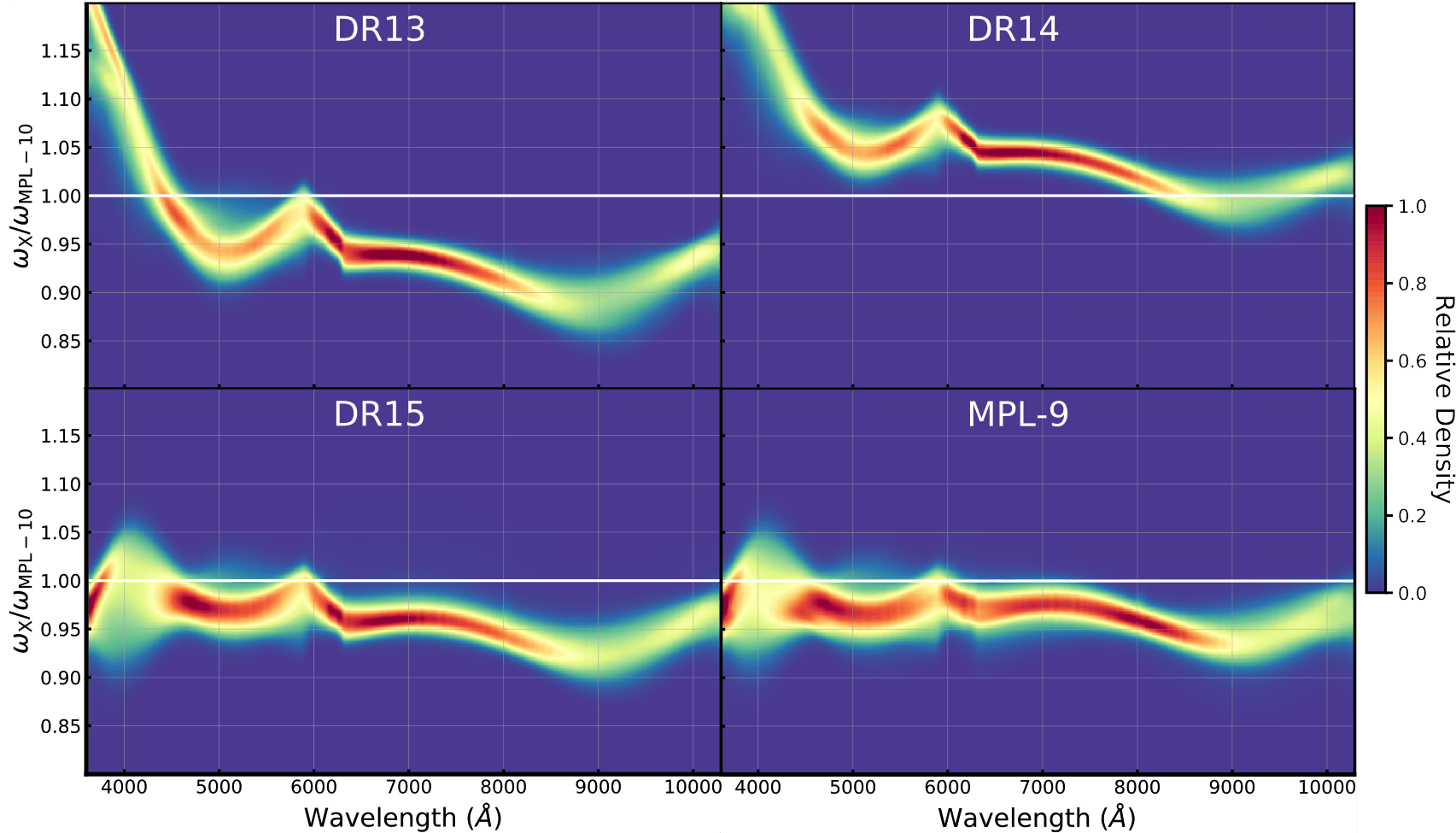}
\end{center}
\caption{Evolution of the MaNGA LSF estimate over the lifetime of the survey.  Each panel shows a 
density plot of the ratio between the LSF vectors for all fiber spectra in common between MPL-10 and prior data releases
DR13, DR14, DR15, and MPL-9.}
\label{comparedr.fig}
\end{figure*}

As the MaNGA DRP has evolved over the survey, the estimated instrumental LSF too has changed.  Rather than representing significant changes in the data, as outlined in the previous sections this instead reflects our evolving understanding of the instrument and improvements to our methods of characterizing the data.
These changes are summarized in Figure \ref{comparedr.fig}, which plots the ratio between $\omega$ for
all MPL-10 fiber spectra that were included in four of the major internal/external
MaNGA data releases and shows that the MaNGA LSF estimates have generally been converging over the lifetime of the survey.

Relative to MPL-10, MaNGA's initial public data release (DR13, released Summer 2016) systematically overestimated the pre-pixellized
LSF in the far blue by 10-15\% while underestimating it in the red by 5-10\%.  
This changed with the release of DR14 (Summer 2017) which made an initial correction to broaden the pre-pixellized LSF measurements
to post-pixellized values during the skyline adjustment stage, in addition to including a term to  account for the broadening factor
contributed by the wavelength rectification of fiber spectra \citep[see][]{dr14}.
These changes generally improved performance at red wavelengths but were an overcorrection, 
leading to a $\sim 5$\% effective overestimate of the LSF around H$\alpha$.

DR15 (Summer 2018) for the first time provided simultaneous measurements of the pre-pixellized and post-pixellized LSF in full cube format, fixed 
the LSF overestimate in the far blue (by rejecting the partially-blended Cd I $\lambda\lambda 3610$
and low-quality Hg II $\lambda\lambda 3984$ arclamp lines and substituting Hg I $\lambda\lambda 3663$ instead), and implemented
a Gaussian-comb solution for measuring the broadening due to wavelength rectification (instead of relying upon measurements of individual lines).
As illustrated by Figure \ref{comparedr.fig}, the combination of these changes brought all wavelengths relatively well in line with MPL-10, but with a $\sim$ 4\%
underestimate of the LSF around H$\alpha$.  Minor additional changes to polynomial fitting orders and rejection of additional skylines implemented in MPL-9 in
turn decreased this difference to around 2-3\%.

Relative to MPL-9, MPL-10 completely overhauled many aspects of the LSF estimation.  Most fundamentally, instead of using legacy C code 
dating back to the original
SDSS spectro survey to measure the pre-pixellized LSF (and later bootstrap the corresponding post-pixellized LSF), it instead uses new code to measure
the post-pixellized LSF and later bootstraps the pre-pixellized values based on extensive Monte-Carlo simulations (\S \ref{prepost.sec}).
Additionally, MPL-10 introduced further modifications to the reference arc and sky line lists and polynomial fitting orders.

The scientific impact of this evolution in the estimated LSF will depend on the range of 
astrophysical velocity dispersions considered by a given analysis.
Above $\sigma_{\Ha} = 100$ \kms, the systematic error in recovered velocity dispersions after correcting for the instrumental resolution will be less than 3\%.  At velocity dispersions far below
the instrumental resolution though the changes in the estimated LSF become 
increasingly important, and
around $\sigma_{\Ha} = 30$ \kms\ analyses
using DR13, DR14, or DR15 data will have derived
values that are systematically too
high by $\sim$ 30\%, too low by $\sim$ 30\%, and
too high by $\sim$ 20\% respectively compared to MPL-10.

\section{Spectral LSF in the MaNGA DAP}
\label{dap.sec}

The MaNGA Data-Analysis Pipeline (DAP) is the survey-led software package that derives astrophysical measurements from the DRP data cubes.
These measurements are produced by five core modules with the following primary purposes: (1) spatially bin the data to meet a $g$-band continuum SNR using the Voronoi binning method of \citet{vorbin03}; (2) measure stellar kinematics by performing full-spectrum fitting using pPXF \citep{ppxf04, cappellari17} with the emission lines masked; (3) calculate moments of the emission-line profiles to obtain non-parametric (pixel-summed) fluxes and equivalent widths; (4) fit single-component Gaussian profiles to the emission lines with simultaneous readjustment of the stellar continuum using pPXF; and (5) measure spectral indices, including both absorption-line indices \citep[e.g., H$\delta_{\rm A}$;][]{worthey97} and ``break'' indices \citep[e.g., D4000; ][]{d4000}. Detailed descriptions of these modules and tests of the efficacy and performance of the DAP are provided by \citet{westfall19} and \citet{belfiore19}, with the latter focusing specifically on the emission-line modeling.  Below, we discuss the full-spectrum-fitting modules of the DAP (i.e., modules 2 and 4 above) and their corresponding treatment of the spectral LSF.

\subsection{Overview of spectral fitting in the DAP}
\label{emtpl.sec}

The pPXF approach operates on the fundamental assumption that a galaxy spectrum is composed of a linear combination of template spectra convolved with a template-dependent line-of-sight velocity distribution (LOSVD). Ignoring any multiplicative (e.g., attenuation, flux-calibration) or additive (e.g., sky-subtraction) effects and explicitly including the convolution by the 
pre-pixel spectral resolution kernel, $\mathcal{R}(\lambda)$, and the pixel-sampling kernel, $\mathcal{S}(\lambda)$, we can write the underlying pPXF assertion as \citep[cf.][Eqn.~11]{cappellari17}:
\begin{equation}
G_j \equiv G^\prime \ast \mathcal{R}_j \ast \mathcal{S} \approx \sum_i w_i (T^\prime_i \ast \mathcal{L}^\prime_i \ast \mathcal{R}_j \ast \mathcal{S}),
\label{eq:ppxf}
\end{equation}
where $\ast$ denotes a wavelength-dependent convolution, $\mathcal{L}^\prime_i$ is the {\it intrinsic} LOSVD kernel, $G^\prime$ and $T^\prime_i$ are the {\it intrinsic} spectra of the galaxy and template object respectively, and the summation runs over each of the individual template objects.
We have additionally subscripted the observed galaxy spectrum, $G$, and the associated spectral resolution kernel with the index $j$ to emphasize that MaNGA observations have a range in spectral resolution that vary both between observations and between spaxels in a given datacube (Figure \ref{cubelsf.fig}).
Except for templates based on theoretical models, the $T^\prime_i$ are not generally known; instead we have observed spectra that are taken with their own spectral resolution and sampling kernels; i.e., $T_i \equiv T^\prime_i \ast \mathcal{R}_i \ast \mathcal{S}_i$.  Solving for $T^\prime_i$, Equation \ref{eq:ppxf} becomes
\begin{equation}
G_j \approx \sum_i w_i \left\{ T_i \ast \mathcal{L}^\prime_i \ast (\mathcal{R}_j\ \overline{\ast}\ \mathcal{R}_i) \ast (\mathcal{S} \ \overline{\ast}\ \mathcal{S}_i)\right\},
\label{eq:ppxf2}
\end{equation}
where we use $\overline{\ast}$ to signify a wavelength-dependent {\it deconvolution}.
Although pPXF allows for template spectra with a pixel sampling that is a fixed integer factor smaller than the sampling of the galaxy spectrum,\footnote{In this case, the convolution by $(\mathcal{S}\ \overline{\ast}\ \mathcal{S}_i)$ is accomplished via a simple integer rebinning of the convolved template spectra.  This functionality ensures that the LSF of the template spectra is still well sampled, even when the templates have substantially higher spectral resolution compared to the galaxy spectrum.} any differences in the (pre-pixelized) spectral resolution of the templates and the galaxy data is ignored.

Standard practice is to use templates that have been convolved to exactly the same spectral resolution as the galaxy data, either by observing templates with the same instrument setup or by degrading the spectral resolution of a set of templates to match the resolution of the galaxy data.
As we note in \citet{westfall19} however this is hard to do in practice for many reasons: the reproducibility of the instrument setup, the uncertainty in the determination of the wavelength-dependent LSF for both the galaxy and template spectra, and the redshift difference between the galaxy and template objects.  Moreover, an analysis of the relevant error propagation implies that it is almost always better to use templates with higher spectral resolution than the galaxy spectra \citep[][Appendix B]{westfall19}. 
In the MaNGA DAP, we therefore chose to use pPXF to fit the galaxy data using templates at their native spectral resolution.  This means that we must construct a correction that accurately removes the effect of the template-galaxy spectral-resolution difference to recover the parameters of the astrophysical LOSVD.  In terms of Equation \ref{eq:ppxf2}, our approach is to have pPXF fit $\mathcal{L}_{ij} = \mathcal{L}^\prime_i \ast (\mathcal{R}_j\ \overline{\ast}\ \mathcal{R}_i)$ --- the kernel composed of the astrophysical LOSVD convolved with the template-galaxy instrumental resolution difference --- and then we construct a correction to $\mathcal{L}_{ij}$ that yields $\mathcal{L}^\prime_i$.  The details of how the DAP constructs the corrections are different for the ionized gas and the stars;
however, both currently assume that the template LSF, the MaNGA LSF, and the LOSVD are single-component Gaussian profiles.  This means that the only parameter of $\mathcal{L}_{ij}$ that requires correction is the velocity dispersion.

The key difference between the velocity-dispersion corrections derived for the stars and ionized gas is that the former ignores the wavelength dependence of $(\mathcal{R}_j\ \overline{\ast}\ \mathcal{R}_i)$, whereas the latter is effectively treated on a line-by-line, wavelength-dependent basis.

The construction of the emission-line templates is described in detail by \citet[][Section 9.1]{westfall19}.  Taking advantage of the analytic Fourier transform of a Gaussian line profile \citep{cappellari17}, we set the spectral resolution of the emission-line templates to match the resolution of the MaNGA datacube up to a quadrature offset in $\omega_{\rm PRE}$, to first order.  This is only done once per datacube, meaning that there are second-order differences between the resolution of the emission-line templates and the data given the spaxel-to-spaxel variations in the LSF and the spectral variation in the LSF over the velocity scale of the galaxy's internal motions.  That is, $(\mathcal{R}_j\ \overline{\ast}\ \mathcal{R}_i)$ is nearly constant for the emission-line templates.  Regardless, the DAP fits the velocity dispersion of each line (except for the line doublets listed in Section \ref{tied.sec}) independently,
which allows pPXF to account for the second-order LSF effects during the fit.  Moreover, by adding the template-line velocity dispersion to the pPXF measurement in quadrature, the measurement reported by the DAP for each emission line is exactly the pre-pixelized velocity dispersion of the {\it observed} line profile ($\mathcal{L}^\prime_i \ast \mathcal{R}_j$ from Equation \ref{eq:ppxf}).\footnote{Tests have shown that this approach provides results that are virtually identical to a direct fit of a pixelized Gaussian profile to each line profile.}  The corrections needed (and provided by the DAP) to calculate the velocity dispersion of $\mathcal{L}^\prime_i$ is, therefore, 
the {\it pre-pixelized} width, $\omega_{\rm PRE}$, of the instrumental line profile ($\mathcal{R}_j$ from Equation \ref{eq:ppxf}) at exactly the best-fitting centroid of the line. Specifically, the corrected velocity dispersion of, e.g., the H$\alpha$ line is:
\begin{equation}
\sigha = \sqrt{\sigma^2_{\Ha,{\rm obs}} - \omega^2_{\rm PRE}},
\label{emission.eqn}
\end{equation}
where $\sigma_{\Ha,{\rm obs}}$ is the pre-pixelized velocity dispersion of the observed line.

In DR15 and subsequent MPLs, the templates used to measure stellar kinematics are based on a
set of 42 composite spectra generated by the hierarchical clustering of the 985
empirical stellar spectral of the MILES library \citep{miles1,falconbarroso11}: i.e., the ``MILES-HC'' library \citep[][Section 5]{westfall19}.  The MaNGA spectra are fit only over the MILES spectral range ($3575 {\rm \AA} < \lambda_{\rm rest} < 7400 {\rm \AA}$).  We adopt a wavelength-independent instrumental resolution of $\Delta\lambda = 2.5$ \AA\ for the MILES library \citep[][cf., \citealt{beifiori11}]{falconbarroso11}.\footnote{We assume that this is the pre-pixelized LSF width, consistent with our results in Section \ref{shravan.sec}.  We note the subtle difference between the value used by the DAP from \citet{falconbarroso11} and the value of $\Delta\lambda = 2.54$ \AA\ quoted by \citet{beifiori11}.  The difference between these two measurements amounts to a 1.6\% difference in the MILES instrumental FWHM, which leads to a $\sim 3$\% difference in $\sigma_\ast$ around 50 \kms.  However, note that this uncertainty in the MILES spectral resolution has no effect on the ionized-gas velocity dispersions.}  Given the MaNGA resolution from Figure \ref{rsslsf.fig}, $(\mathcal{R}_j\ \overline{\ast}\ \mathcal{R}_i)$ --- and therefore $\mathcal{L}_{ij}$ ---  is wavelength dependent; however, we use pPXF to instead fit a wavelength-independent parametrization of $\mathcal{L}_{ij}$.  Given the complications involved in allowing for a wavelength-dependent $\mathcal{L}_{ij}$ and the fact that the stellar kinematics are determined by {\it all} the absorption features in the MaNGA spectra, we adopt a simple correction for the stellar velocity dispersion derived from the average difference in the pre-pixelized MaNGA and MILES resolution over the fitted spectral range.  The corrected velocity dispersion of the stars is therefore:
\begin{equation}
\sigma_\ast = \sqrt{ \sigma^2_{\ast,{\rm obs}} - \langle\omega^2_{\rm PRE} - \omega^2_{\rm MILES}\rangle_\lambda},
\label{absorption.eqn}
\end{equation}
where $\sigma_{\ast,{\rm obs}}$ is the velocity dispersion of $\mathcal{L}_{ij}$,  $\omega_{\rm MILES}$ is the pre-pixelized instrumental dispersion of the MILES spectra, and the average quadrature difference in the instrumental dispersion ($\langle\omega^2_{\rm PRE} - \omega^2_{\rm MILES}\rangle_\lambda$) is computed over the fitted wavelength range.  This approach is shown to be sufficiently accurate for our purposes \citep[Figure 17 from][see also Section \ref{shravan.sec}]{westfall19}.

\subsection{Updates to the DAP since DR15}

Since DR15 we have made a few key improvements to the DAP compared to the algorithms described by
\citet{westfall19} and \citet{belfiore19}.  Here we update this information briefly before discussing the
reliability of the velocity dispersion measurements in \S \ref{performance.sec}.

\subsubsection{Updated stellar-continuum templates during emission-line modeling}

Measurements of the stellar and emission-line kinematics are performed by separate modules in the DAP (i.e., modules 2 and 4 respectively).  The stellar kinematics are measured first, with the expected locations of any emission lines masked such that only stellar templates are included in the pPXF fit.  For the emission-line kinematics the stellar and emission-line templates are combined for the pPXF fit with the stellar kinematics fixed to the results obtained by the previous module.  An advantage of performing the measurements in separate modules is to, e.g., ensure the stellar kinematics are uncorrelated with and insensitive to the modeling of the gas components.  Although the stellar kinematics are held fixed during the emission-line modeling, the relative weights of the stellar-continuum templates are re-optimized jointly with the emission-line templates to ensure the emission-line properties are not biased by the previous fit \citep[see][Section 5.2]{belfiore19}.  This is particularly important when the DAP is used in its ``hybrid binning'' mode, where the spatial bins used to determine the stellar kinematics ($g$-band S/N $\gtrsim 10$) are deconstructed and the emission-line parameters are fit per spaxel \citep[][Section 9.2]{westfall19}.

In DR15, both full-spectrum fitting modules used the MILES-HC templates to model the stellar continuum.  However, given that the templates are re-optimized, this is not strictly required
\citep[see, e.g., discussion by][Section 4]{belfiore19}.
Therefore, for MPL-9 and later, we switch from the MILES-HC templates in the stellar-kinematics module to a set of templates derived from our own stellar template library \citep[MaStar;][]{yan19} in the emission-line module.  This template switch allows us to continue to leverage the higher resolution of the MILES-HC spectra for the stellar kinematics, while taking advantage of the longer spectral range of the MaStar spectra to allow fits to
lines such as [\ion{S}{3}]$\lambda\lambda9071,9533$ lines.
Although the details are still being explored, we expect the use of the MaStar spectra to be subject to systematic uncertainties that are of the same order as those found by \citet[][Section 4]{belfiore19}.

\subsubsection{Updated line list and tied parameters}
\label{tied.sec}

Since DR15 we have fit additional emission lines to the MaNGA data beyond those listed by
\citet[][Table 3]{westfall19} and \citet[][Table 1]{belfiore19}, in part to 
take advantage of the spectral range of the MaStar templates.

These additional lines include the Balmer lines H12~$\lambda3750$ and  H11~$\lambda3771$,  \ion{He}{1}~$\lambda3889$, [\ion{N}{1}]$\lambda\lambda$5198,5200, \ion{He}{1}~$\lambda7065$, [\ion{Ar}{3}]$\lambda\lambda$7136,7751, Pa-$\eta~\lambda$9015, Pa-$\zeta~\lambda$9229, and $[$\ion{S}{3}$]\lambda\lambda$9069,9531.  We have also changed the adopted fixed flux ratios of some doublets based on an improved calculation (see Table \ref{emline.table}).

\begin{deluxetable}{lr}
\tablecolumns{2}
\tablewidth{0pc}
\tabletypesize{\scriptsize}
\tablecaption{Tied Emission Lines\tablenotemark{a}}
\tablehead{\colhead{Doublet/Pair} & \colhead{Flux Ratio}}
\startdata
$[$\ion{O}{2}$]\lambda\lambda$3726,3729 & \nodata \\[2pt]
$[$\ion{Ne}{3}$]\lambda\lambda$3869,3967 & $1/0.3$ \\[2pt]
\ion{He}{1}$~\lambda$3889, H$\zeta$ & \nodata \\[2pt]
$[$\ion{O}{3}$]\lambda\lambda$4959,5007 & 0.35 \\[2pt]
$[$\ion{N}{1}$]\lambda\lambda$5198,5200 & \nodata \\[2pt]
$[$\ion{O}{1}$]\lambda\lambda$6300,6364 & $1/0.32$ \\[2pt]
$[$\ion{N}{2}$]\lambda\lambda$6548,6583 & 0.34 \\[2pt]
$[$\ion{S}{3}$]\lambda\lambda$9069,9531 & 0.41
\enddata
\tablenotetext{a}{All pairs listed have their velocity dispersions tied.  Only those lines with data in the second column have their fluxes tied.}
\label{emline.table}
\end{deluxetable}

While the MaNGA DAP treats the stellar component as a single kinematic component \citep[cf.,][]{tabor2019, shettyad},
each emission line is allowed to be largely independent with a few exceptions.
First, the redshift is forced to be the same for all emission lines in a given spaxel by tying all the velocities together, which helps stabilize the fit (particularly for lines with relatively low flux) against results biased by noise. Second, virtually all of the
35 velocity dispersions are left free, allowing us to mitigate systematic errors due to the second-order wavelength dependence of $(\mathcal{R}_j\ \overline{\ast}\ \mathcal{R}_i)$ for the emission-line templates.
The only exceptions to this are the 8 line doublets listed in Table \ref{emline.table}; note that \ion{He}{1}$~\lambda$3888.6 and H$\zeta ~\lambda$3889.1 are tied as a practical matter, given that they are unresolved by MaNGA.


\subsection{Precision and accuracy of the velocity-dispersion measurements}
\label{performance.sec}

Using a combination of idealized recovery simulations and analysis of repeat observations (i.e., galaxies observed on more than one plate and processed into independent data cubes), in \citet[][Section 7.5]{westfall19} 
and \citet[][Section 3]{belfiore19} we explored the precision and accuracy of the observed line width measurements $\sigma_0$ for the stars and emission lines respectively.
Specifically, we found that the formal uncertainties $\varepsilon_0$ 
reported by the DAP\footnote{I.e., the
errors determined from the fit covariance matrix \citep[see, e.g.,][Section 15.5]{NumRecThirdEd} and the
inverse variance vectors provided by the DRP.}
were generally reliable based on comparison with repeated observations, albeit somewhat underestimated
at SNR $> 100$ due to small uncertainties in the astrometric registration of individual MaNGA exposures
\citep[see discussion by][]{belfiore19}.

However, neither \citet{westfall19} nor \citet{belfiore19} explored the propagation of these 
uncertainties in the observed line widths (along with uncertainties in the estimated spectral LSF) to the
effective uncertainties in the underlying astrophysical LOSVD.
Here, we explore the influence of the LSF measurements on the accuracy of the emission-line and absorption-line velocity dispersions, and specifically target the accuracy of astrophysical measurements of the LOSVD $\sigha$ (Equation \ref{emission.eqn}) and $\sigma_\ast$ (Equation \ref{absorption.eqn}). The uncertainty $\varepsilon_{\rm rec}$ of these astrophysical widths will be a strong function of both the total SNR and the astrophysical width itself (as measurements will become less reliable far below the instrumental resolution).


\begin{figure}
\begin{center}
\includegraphics[width=\columnwidth]{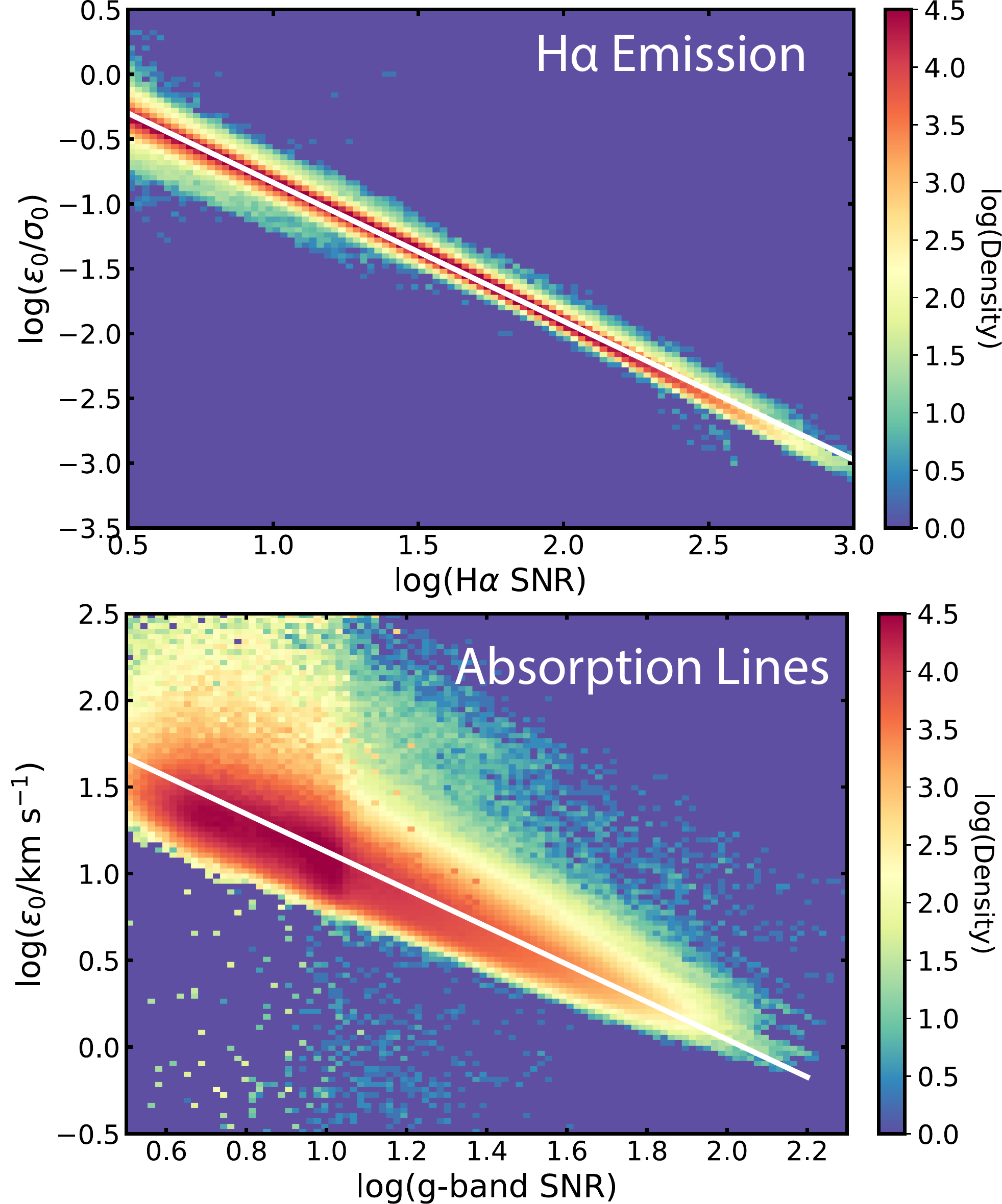}
\epsscale{1.1}
\end{center}
\caption{Top panel: Fractional uncertainty $\varepsilon_0/\sigma_0$ in the observed emission line width reported by the DAP as a function of emission line SNR (colored density map).  Bottom panel: Uncertainty $\varepsilon_0$ in the stellar absorption line measurements as a function of the $g$-band continuum SNR.  In both panels the white solid line indicates our functional fit to the relation.  Note that the sharp feature around SNR $= 10$ for the stellar absorption line measurements is an artifact of the Voronoi binning.}
\label{snr_err.fig}
\end{figure}

\begin{figure*}
\begin{center}
\includegraphics[width=\textwidth]{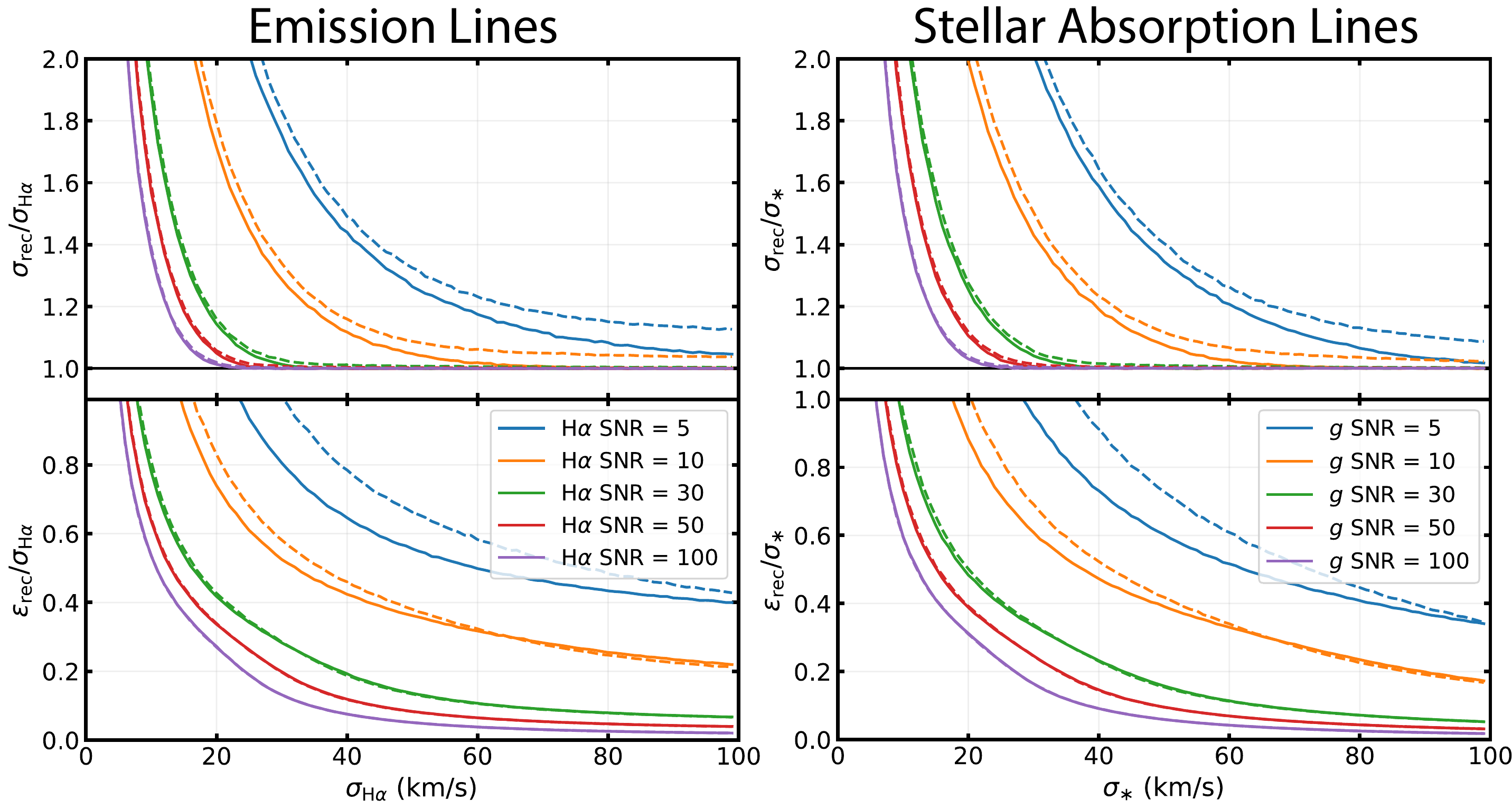}
\epsscale{1.1}
\end{center}
\caption{Top panels: Sigma-clipped mean recovered velocity dispersion $\sigma_{\rm rec}$
as a function of the intrinsic astrophysical velocity dispersion for Monte Carlo simulations of emission lines and stellar absorption lines at a variety of SNR.  Recovered velocity dispersions $\sigma_{\rm rec}$ are systematically larger than the astrophysical velocity dispersions ($\sigma_{\Ha}$ and $\sigma_{\ast}$) at low dispersion due to the preferential loss of spaxels with observed line widths below the instrumental resolution from the sample.  Bottom panels: Typical fractional RMS error $\varepsilon_{\rm rec}/\sigma_{{\rm H}\alpha}$
and $\varepsilon_{\rm rec}/\sigma_{\ast}$ as a function of
intrinsic dispersion for Monte Carlo simulations at a variety of SNR.  In all panels dashed lines represent simulations that adopt an inverse gamma error distribution, while solid lines represent simulations that adopt a simiplified gaussian distribution.}
\label{uncertainty.fig}
\end{figure*}

In Figure \ref{snr_err.fig}, we plot the error in the observed velocity dispersion as a function of the SNR 
for both the H$\alpha$ line and the stellar continuum.
For the H$\alpha$ line, there is a very tight correlation between the fractional error and
the SNR --- i.e., $\varepsilon_0 \propto \sigma_0 ({\rm S/N})^{-1}$ --- as expected when fitting a Gaussian line profile.  The distribution for $\sigma_\ast$ is more complicated because $\varepsilon_0$ becomes increasingly independent of $\sigma_0$ as $\sigma_0$ becomes small relative to the instrumental resolution; this effect is illustrated by analysis of both idealized simulations and repeat observations by \citet[][Figures 19 and 20]{westfall19}.

We use the tight relationship between the fractional error in $\varepsilon_0/\sigma_0$ and SNR for the emission lines to explore the precision and accuracy of the astrophysical measurements of $\sigma_{\Ha}$ using a series of Monte Carlo simulations.
The input for each simulation is the astrophysical LOSVD $\sigma_{\Ha}$ (for which we adopt
a grid from 1-100 \kms\ stepped every 1 \kms), the emission line SNR (for which we adopt
 a range from SNR = 5 -100), the instrumental LSF $\omega_{\rm PRE}$ (which for simplicity we fix to be 67.6 \kms, the median at the wavelength of H$\alpha$ for
the MPL-10 sample), 
the fractional error in the observed line width $\varepsilon_0/\sigma_0$ at a given SNR from the relation shown in Figure \ref{snr_err.fig}, 
and an estimated 3\% statistical uncertainty in the LSF
(see Section \ref{diskmass.sec}).
For each combination of $\sigma_{\Ha}$ and SNR we draw $10^5$ samples from the error distributions\footnote{We test both an inverse gamma distribution 
\citep[][Section 24.1]{MacKay03} and a simplified gaussian distribution model with matched
mode and rms.  The differences between these distributions are negligible for high SNR, but
become appreciable when the fractional error in $\sigma_0$ is $\gtrsim 10$\%.}
for both $\sigma_0$ and $\omega_{\rm PRE}$, 
and compute the mean ($\sigma_{\rm rec}$) and rms ($\varepsilon_{\rm rec}$)
of the $10^5$ astrophysical velocity dispersions recovered
following Equation \ref{emission.eqn}.

The stellar velocity dispersions are significantly more complicated to model in detail as they are derived from a simultaneous fit across a wide range of different wavelengths.  However, we obtain a rough estimate of their reliability by performing a similar series of Monte Carlo simulations using $\omega_{\rm PRE} = 74$ \kms\ (i.e., an average value throughout the wavelength range of interest), combined with the observed error distribution in the observed stellar line widths.

As illustrated by Figure \ref{uncertainty.fig}, the recovered line widths are most reliable at high SNR and high intrinsic astrophysical velocity dispersions, and the effective errors $\varepsilon_{\rm rec}$ in the recovered line width increase dramatically towards lower SNR.  In addition, below the instrumental resolution we note a systematic 
positive bias in the recovered velocity dispersions whose strength increases towards
lower SNR and lower $\sigma_{\Ha}$ or $\sigma_{\ast}$. This expected behaviour arises because of the asymmetric error distribution; namely, data points whose measured line widths (after application of mock measurement errors) are less than the instrumental line width produce imaginary astrophysical widths following Eqns. \ref{emission.eqn} or \ref{absorption.eqn} and are thus preferentially lost from the sample.

In a per-spectrum sense Figure \ref{uncertainty.fig} can be interpreted as giving the SNR cut required in order for the measurements to reach a given accuracy.  In order to obtain velocity dispersions at $\sigha = 20$ \kms\ for which all data points have less than 1 \kms\ systematic error for instance, the spaxel sample must be restricted to those with H$\alpha$ SNR $> 50$, for which the typical statistical uncertainty $\varepsilon_{\rm rec}$ will be about 5 \kms.  Although more stringent cuts in SNR would produce samples with less systematic bias, the gain in such cases must be weighed against the dramatically decreased sample size
at larger SNR thresholds.

We note that similar analyses have been performed for both the SAMI and CALIFA
surveys, for which qualitatively similar trends are observed.
\cite{fb17}, for instance, note that the
recovered stellar velocity dispersions in CALIFA are 
systematically larger than expected below $\sigma_{\ast} = 40$ \kms, at which
point the random typical uncertainty in individual measurements is about 20\%.  Likewise,
\citet{fogarty15} and \citet{vds17} find that systematic uncertainties in the instrumental resolution dominate the SAMI error budget for stellar velocity dispersions 
below $\sigma_{\ast} = 35$ \kms\ 
and recommend a variety of quality cuts in both SNR and $\sigma_{\ast}$ accordingly.


\subsection{Consistency between Multiple Lines}
\label{multlines.sec}

\begin{figure}
\begin{center}
\includegraphics[width=\columnwidth]{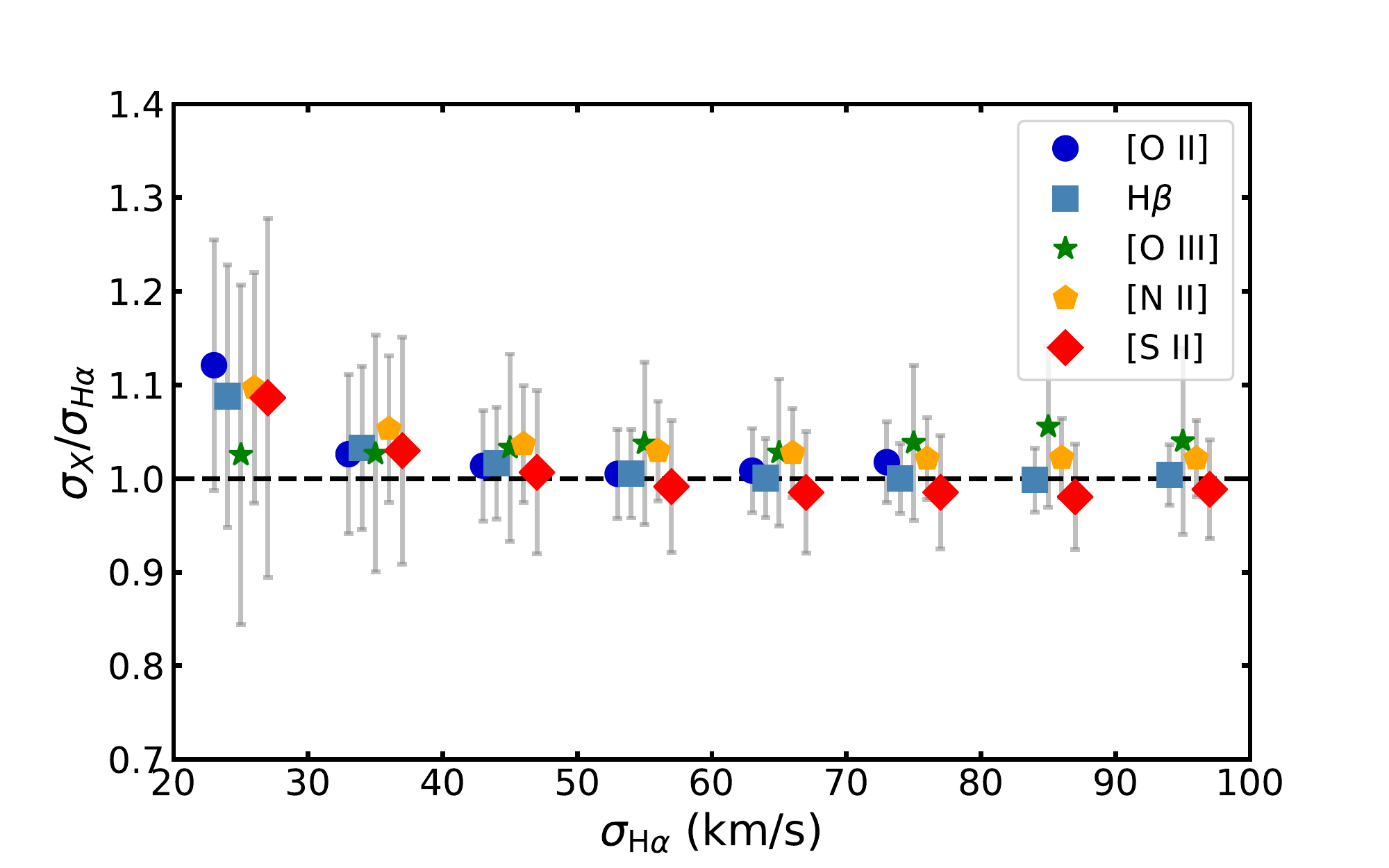}
\end{center}
\caption{Ratio between H$\alpha$ velocity dispersion and velocity dispersions of \otwo\ $\lambda 3727$, \Hb\, \othree\ $\lambda 5008$,
\ntwo\ $\lambda 6584$, and \stwo\ $\lambda 6717$
as a function of $\sigma_{\Ha}$.  For each emission line the sample is limited to star-forming spaxels for which 
both \Ha\ and the other emission line are detected with SNR $> 50$.  Filled points show the
sigma-clipped mean of the distribution while the
error bars indicate the $1\sigma$ width of the observed distribution (uncertainties in the mean
are smaller than the symbols).
}
\label{emlines.fig}
\end{figure}

In \citet[][see their Fig. 21]{belfiore19} we noted that the astrophysical velocity
dispersions computed from a variety of nebular emission lines were broadly consistent with those estimated using \Ha.
In Figure \ref{emlines.fig} we repeat this exercise for the MPL-10 data products
and plot LOSVD ratios as a function of $\sigma_{\Ha}$ for all star-forming 
spaxels\footnote{Defined here as those with \othree/\Hb\ vs \ntwo/\Ha\ line ratios
below the relation defined by \citet{kauffmann03}; see further discussion in Law et al. ({\it in prep}).}
in which both emission line are detected with SNR $> 50$.
Given the large wavelength difference between \otwo\ and \Ha, we additionally restrict 
\otwo\ observations to those with a Balmer decrement indicative of minimal dust attenuation ($f_{\Ha}/f_{\Hb} < 3.5$).

We find that \otwo, \Hb, \othree, \ntwo, and \stwo\ velocity dispersions all match
$\sigma_{\Ha}$ to within a few percent, suggesting that there are no significant wavelength-dependent
errors in the MaNGA LSF compared to the performance around \Ha.
Indeed, for $\sigma_{\Ha} > 30$ \kms\ the small offsets that we see between the dispersions of
different ions may be genuinely astrophysical in origin as the relative offset appears to be correlated
with the ionization energy.
The S$^+$ ion for instance has an ionization energy of 10.4 eV, and median $\sigma_{\stwo}/\sigma_{\Ha} = 0.989$, while the H$^+$ and O$^+$ ions (both with 13.6 eV ionization energy)
have median $\sigma_{\Hb}/\sigma_{\Ha} = 1.004$ and $\sigma_{\otwo}/\sigma_{\Ha} = 1.014$ respectively.
Likewise, the N$^+$ and O$^{++}$ ions have ionization energies of 14.5 and 35.2 eV respectively, and 
median $\sigma_{\ntwo}/\sigma_{\Ha} = 1.027$ and $\sigma_{\othree}/\sigma_{\Ha} = 1.038$.

We note, however, that all four lines except \othree\ show a $\sim$ 10\% turnup in their velocity dispersion
relative to \Ha\ at the lowest values of $\sigma_{\Ha} \sim 20$ \kms.  If this were due to a systematic
error in the LSF, it would imply a $\sim$ 0.8\% offset relative to \Ha.  However, this appears implausible given the small separation between \Ha\ and \ntwo\ in wavelength, and the well-sampled, well-behaved LSF
in this range (see, e.g, Figure \ref{rsslsf.fig}).  Likewise, interpolation of the positivity bias
from Figure \ref{uncertainty.fig} due to the differential SNR of \Ha\ and the other emission lines suggests
that this could account for at most a 1\% deviation rather than the 10\% observed.

The few-percent trends visible in Figure \ref{emlines.fig} may therefore be telling us more about the ionization structure
of H II regions in the MaNGA galaxy sample rather than any low-level, systematic, and wavelength-dependent
uncertainty in the spectral LSF of the instrument.
Indeed, as we demonstrate in a forthcoming contribution (Law et al., {\it in prep}) typical gas-phase
velocity dispersions can vary substantially depending on the selection method since these kinematics
are strongly correlated with ionization mechanism as traced by line ratios
such as \ntwo/\Ha, \stwo/\Ha, and \othree/\Hb.

\section{Beam Smearing}
\label{beamsmear.sec}

An additional consideration in the use of any velocity dispersions provided by the MaNGA DAP is the impact of beam smearing, 
or the effective broadening of velocity
dispersions in a given spaxel by velocity gradients in the galaxy on scales comparable to the MaNGA point spread function.  
The DAP does not correct for this effect, but
given the typical MaNGA spatial resolution of 
2.5 arcsec FWHM \citep[see, e.g., Fig. 17 of][]{law16} beam smearing can be significant, especially for edge-on high-mass galaxies at the high redshift end of
the sample.

This problem is well-known in the literature \citep[e.g.,][]{weiner06,epinat10,stott16,johnson18,varidel19} and a variety of techniques have been developed to attempt to correct for it
ranging in sophistication from simply
ignoring the most-affected spaxels \citep[e.g.,][]{zhou17} to quadrature subtraction of the local 
velocity gradient \citep{varidel16,oa18}, Bayesian inference
modeling \citep{varidel19}, and 
dynamical \citep[e.g.,][]{cappellari08} or
3d forward modeling of the observed data \citep[e.g.,][]{bouche15, bekiaris16, dt15}.
In our present study we are bounded by our desire to make as robust a correction as possible, minimize loss of data,
and also use a technique that can be practically applied to all $\sim 10,000$ galaxies in the MaNGA sample without requiring substantial
computing resources.\footnote{At 450 hours per galaxy for instance, applying the BLOBBY3D bayesian mixture algorithm described by
\citet{varidel19} to our data would require roughly 20 times the computing time required to generate the full DRP+DAP survey results
from the raw observational data for MPL-10.}
We therefore adopt a hybrid approach in which we correct for beam smearing estimated from a 3D model based on the observed velocity field
of each galaxy, and additionally mask out from our analysis all spaxels within 4 arcsec radius of the center of each galaxy for which the
beam smearing correction will be most uncertain.

For each MaNGA galaxy, we first mask out all spaxels in the DAP \Ha\ velocity map 
that have H$\alpha$ SNR $< 3$, or that have non-zero data quality bits set in the flux, velocity, or velocity dispersion mask extensions.
Next, we create a 3D model cube matched to the galaxy spaxels in which each non-masked spaxel has a spectrum composed of a 
single emission line normalized to unity 
with a $1\sigma$ width 40 \kms\ and central wavelength shifted by the doppler velocity given by the DAP velocity map.
\footnote{The choice of 40 \kms\ is unimportant, and chosen to be small enough that broadening effects are easy to measure without
producing double-peaked line profiles; the resulting beam smearing correction changes by just $< 1$ \kms\ on average for any choice of model widths from 20-70 \kms.  Likewise, the derived correction is insensitive to whether or not we rebin the velocity field to a smaller pixel scale prior to constructing the 3D model.}
This data cube is spatially convolved with the effective $r$-band PSF of the galaxy, effectively smearing together 
individual spectra in a manner that mimics
the observed beam smearing. 
The resulting spectrum in each spaxel of the convolved cube is then fit with a Gaussian, and 
the initial 40 \kms\ line width
subtracted in quadrature from the measured width in order to determine a map of the effective beam smearing.

\begin{figure*}
\begin{center}
\includegraphics[width=\textwidth]{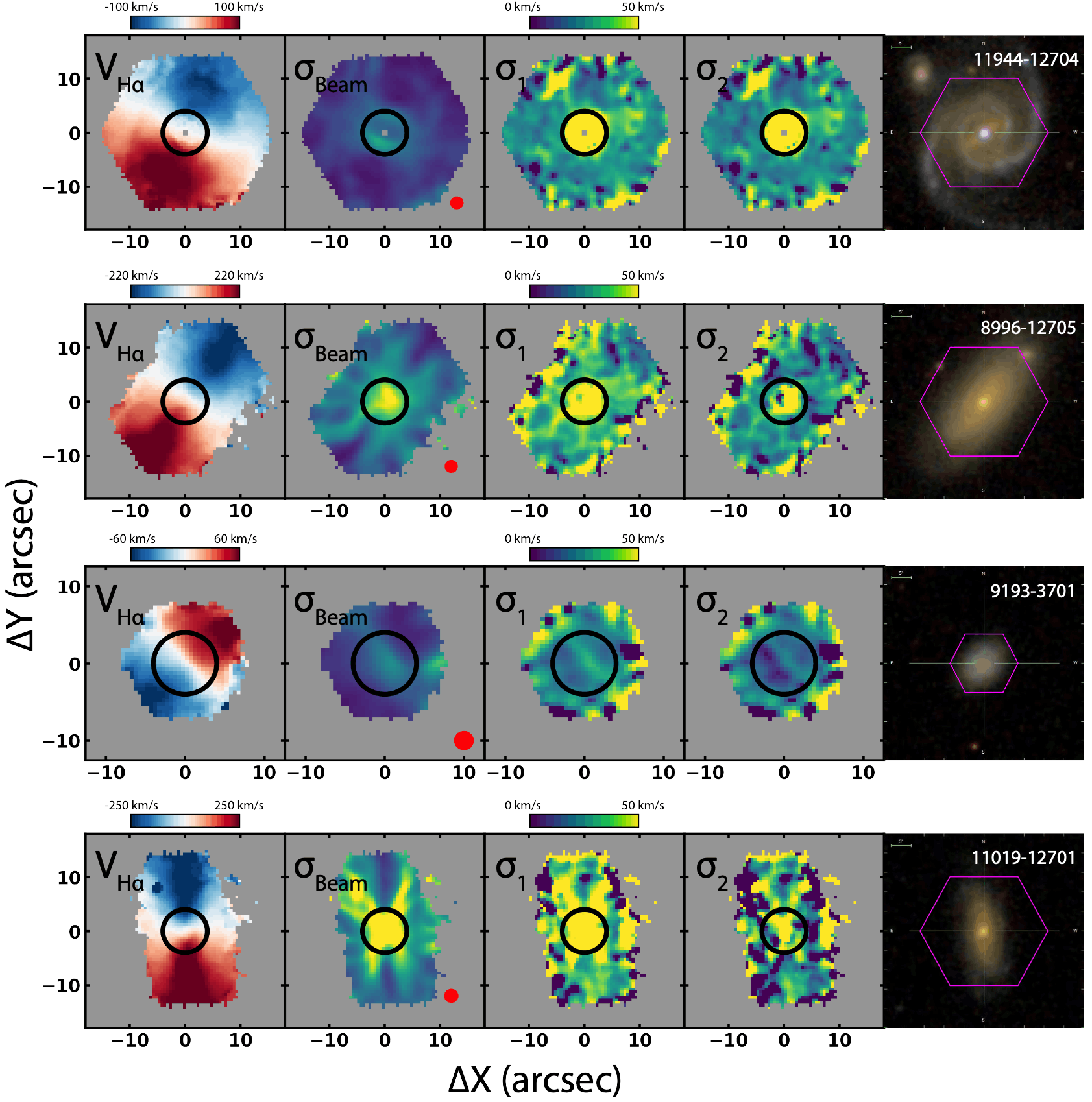}
\end{center}
\caption{Beam smearing correction for four example galaxies selected to be widely representative of the MaNGA star forming galaxy sample.  
Left to right, each panel shows the DAP H$\alpha$ velocity map, our derived beam smearing correction $\sigma_{\rm beam}$,
the gas velocity dispersion before ($\sigma_1$) and after ($\sigma_2$) application of the beam smearing correction, and a color image of the 
galaxy from SDSS imaging with the hexagonal IFU footprint overlaid in purple.  The red circle in the second panel from the left indicates the FWHM
of the MaNGA data, while the solid black circles in the four left-most panels illustrates
the region with radius 4 arcsec that we exclude from our future dispersion analyses.
Examples are shown for 
a high-mass face-on galaxy  (11944-12704; log($M_{\ast}/M_{\odot}$) = 11.1, $i = 36^{\circ}$, $z = 0.069$, $1.5 \, R_{\rm e}$ sample),
a high-mass edge-on galaxy (8996-12705; log($M_{\ast}/M_{\odot}$) = 10.8, $i = 64^{\circ}$, $z = 0.048$, $1.5 \, R_{\rm e}$ sample),
a low-mass face-on galaxy (9193-3701; log($M_{\ast}/M_{\odot}$) = 9.2, $i = 41^{\circ}$, $z = 0.023$, $1.5 \, R_{\rm e}$ sample),
and a distant high-mass edge-on galaxy (11019-12701; log($M_{\ast}/M_{\odot}$) = 11.1, $i = 71^{\circ}$, $z = 0.12$, $2.5 \, R_{\rm e}$ sample).
}
\label{beamsmear.fig}
\end{figure*}

We illustrate this process in Figure \ref{beamsmear.fig}, showing the observed velocity field and velocity dispersion corrections for four
example galaxies that span the range of MaNGA observations from nearby face-on objects (a best-case scenario) to distant
edge-on objects (a worst-case scenario).
As expected, beam smearing corrections vary significantly from $\sim 5-10$ \kms\ at large radii for face-on galaxies such as 11944-12704
to 50 \kms\ or more in the centers of highly inclined galaxies such as 8996-12705.  As illustrated by Figure \ref{beamsmear.fig}, corrected
velocity dispersion maps in which the beam smearing contribution has been
subtracted in quadrature from the observed values are relatively constant outside the central regions of MaNGA galaxies (peaking again in low-SNR 
regions toward the edges of the IFUs).

In addition to astrophysical processes (e.g., AGN and spheroidal popluations of old stars) that can produce broadened values of $\sigma_{\Ha}$ in the central regions of our galaxies
these central peaks may also be due in part to limitations in our beam smearing correction.  Strictly, in the approach described above the 
true galaxy velocity field has been convolved with the observed PSF twice; once
to produce the observed velocity map, and again during the model cube convolution.  This effectively produces shallower 
velocity gradients in central regions of the galaxy, causing us to underestimate the true beam smearing in these regions.  If we instead use a forward model \citep[similar to that used by][]{westfall14} to
simultaneously fit the gas velocity and velocity dispersion fields for the 
example galaxies shown in Figure \ref{beamsmear.fig} we derive beam smearing corrections
for which the median absolute difference is just 2 \kms\ (i.e., insignificant when
subtracted in quadrature), suggesting that outside the central regions the difference
between using the two models is small.\footnote{Indeed, while the simple correction
from the observed velocity map is imperfect in regions where the velocity gradient
is extremely steep, it can nonetheless capture non-rotational kinematic effects missed
by method that assume a fixed intrinsic form for the underlying velocity field.}
It may nonetheless be advisable for 
some science analyses to exclude spaxels within a 3-4 arcsec radius (i.e., $\sim$1.5 times the typical PSF FWHM; black circles in Figure \ref{beamsmear.fig}) from consideration.


The effective beam smearing correction appropriate for stellar kinematics is more complicated to derive in detail, since the stellar kinematics provided by the DAP are derived from a simultaneous fit to multiple absorption lines at different wavelengths.
However, we obtain a first-order estimate by repeating the same analysis as above and simply using the stellar velocity field in place of the ionized gas velocity.


\section{External Assessment of the MaNGA Kinematics}
\label{external.sec}


\subsection{Comparison to high resolution stellar spectra}
\label{xshooter.sec}


\begin{figure*}
\begin{center}
\includegraphics[width=\textwidth]{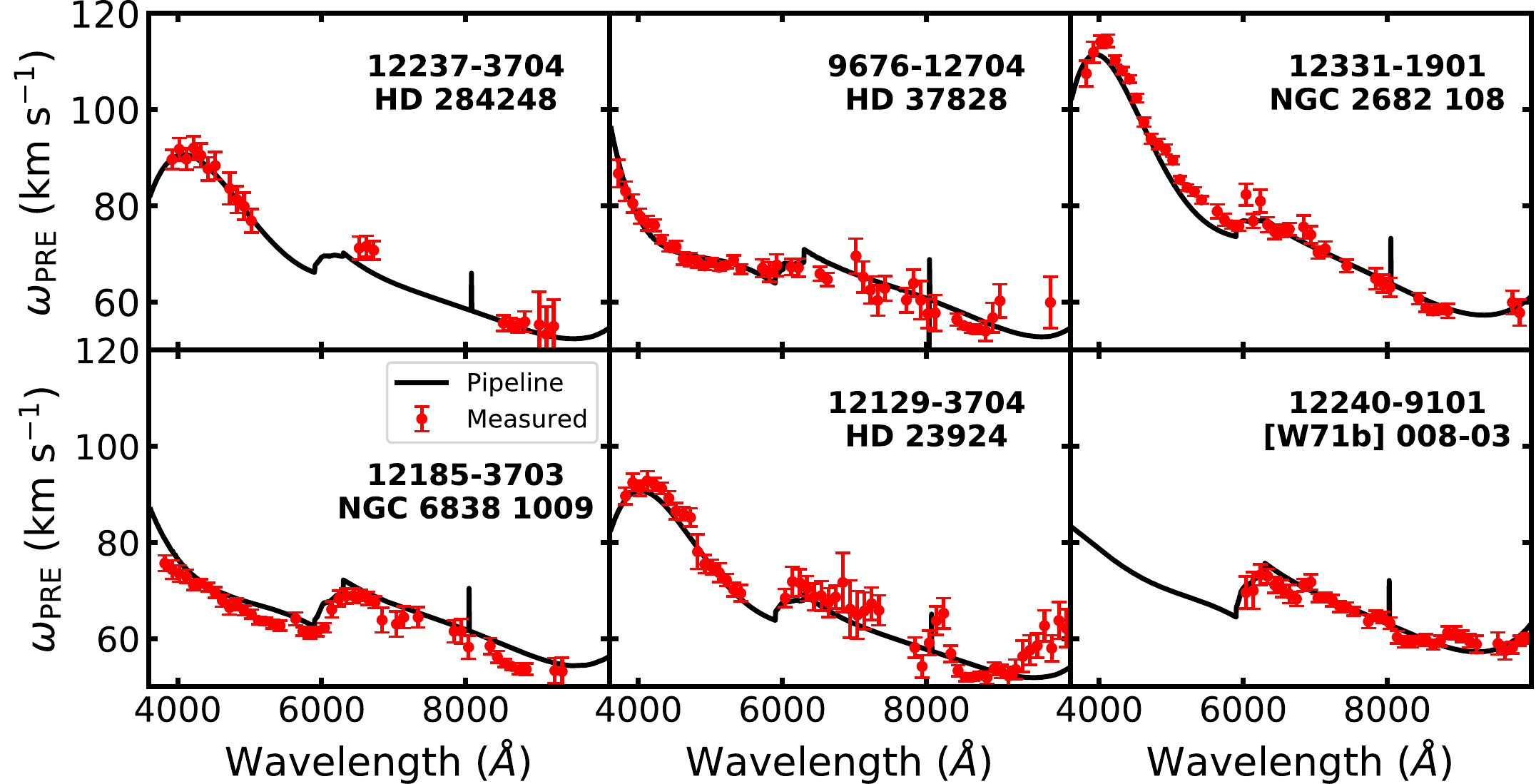}
\end{center}
\caption{Internal pre-pixel DRP estimate of the spectral LSF for six bright MaStar targets (solid black lines) compared to 
empirical estimates derived via comparison of the MaStar spectra with previous $R = 7000 - 11,000$ X-Shooter spectra
(red points, with $1\sigma$ uncertainties).  Points are shown only for
spectral bins in which the MaStar and convolved X-Shooter spectra are
visibly well matched.  No high-quality X-shooter spectra are available for short wavelengths in [W71b] 008-03.
}
\label{xshooter.fig}
\end{figure*}

As a part of the MaStar stellar library program we observed 
six bright stars (HD 284248, HD 37828, NGC 2682 108, NGC 6838 1009, HD 23924, and [W71b] 008-03) that had previously been observed at higher
spectral resolution by X-Shooter.  Since these stars have visual magnitudes
 $g = 6-13$ that are significantly
brighter than typical MaNGA/MaStar targets, observations were made using
custom 10-250 second exposures instead of the usual 900 second exposures.
Since night sky lines are too faint to be observed
reliably in such short exposures, the DRP skips the skyline adjustment step (\S \ref{sky.sec})
 for these observations and relies entirely upon the arc lamp LSF solution.

We compared the final 
one-dimensional spectra of these six stars produced by the DRP to high-resolution template
spectra drawn from the X-Shooter Spectral Library \citep[XSL,][]{chen14}.
Spectra for HD 284248 and HD 37828 were taken from XSL DR1 \citep{chen14} 
and have been processed with a unified resolution of $R = 7000$ across the wavelength range $\lambda = 3000 - 10185$ \AA, while the remaining four stars were taken from
XSL DR2 \citep{gonneau20} and have spectral resolution $R = 9793/11573$ in the MaNGA
blue/red wavelength ranges respectively.
After converting the X-Shooter templates to the MaStar vacuum
rest-frame wavelength solution and rebinning them to a constant pixel size of 30 \kms, we broke the spectra into 200-400 \AA\ windows
stepped every 100 \AA\ and measured the effective spectral resolution based on a convolution of the high-resolution templates.
After rejecting wavelength regions for which the MaStar and X-Shooter spectra do not match each other well (e.g., due to differences in the correction
from telluric absorption bands) we plot the X-Shooter derived LSF of the MaStar spectra against the DRP estimates in Figure \ref{xshooter.fig}. 

As indicated by Figure \ref{xshooter.fig}, the effective spectral LSF derived from the X-Shooter spectra
is generally in excellent agreement with the DRP-estimated LSF throughout the entire
wavelength range.  Although there are small systematic deviations
for some of the stars (e.g., 12185-3703), these are generally within
the uncertainty of the convolution technique given calibration differences
between the spectra.




\subsection{Comparison to the MILES spectral library}
\label{shravan.sec}

While \S \ref{xshooter.sec} qualitatively suggested that the MaNGA LSF estimate was reasonable in the case of a single object with extant
high-resolution data, the MaSTAR  stellar spectral library provides an opportunity to test the pipeline-estimated LSF in a statistical manner
as well.  Barring unusual broadening due to stellar rotation and atmospheric features,  the observed absorption line width in a given 
MaSTAR spectrum should be similar to other stars of similar spectral type.
Hence, comparing these MaSTAR spectra to those from a spectral template set with an established LSF we can test the robustness
of the MaNGA LSF measurements. 

For this test we conduct full-spectrum fitting of a large sample of MaSTAR spectra using the {\tt pPXF} algorithm \citep{cappellari17} and 
the MILES-HC spectral template set
\citep[see \S \ref{emtpl.sec}, and section 5 of][]{westfall19}.
 Since the LSF of the MILES stellar library is both well 
 studied \citep[$\omega_{\rm MILES} = 2.54$ \AA\ FWHM][]{beifiori11} 
 and slightly higher resolution than MaSTAR, a strong test of the accuracy of the MaNGA LSF estimate
 can be provided by comparing the broadening required to reproduce the MaSTAR spectra using the MILES-HC templates:
 \begin{equation}
A = \sqrt{\omega_{\rm Fit}^2 - \omega_{\rm MILES}^2}
\end{equation}
 against the expected broadening factor based on $\omega_{\rm MILES}$ and the pipeline-reported LSF $\omega_{\rm PRE}$:
\begin{equation}
B = \sqrt{\omega_{\rm PRE}^2 - \omega_{\rm MILES}^2}
\end{equation}

In this test we compute $A$ for a random sample of 5,000 spectra from the MaSTAR sample of `Good Visit' spectra, which are single visit observed spectra without extinction issues, having a median S/N per pixel greater than 15 and that pass a visual inspection for other quality problems \citep{yan19}. Due to the limited wavelength coverage of the MILES stellar library (and hence the MILES-HC templates), for this test we fit the MaSTAR spectra within a wavelength range of 3,620-7,400\AA. The full spectrum fit is conducted using 8 additive and multiplicative polynomials in order to account for issues of flux calibration errors, reddening, template mismatch, etc \citep[see, e.g.,][]{westfall19}.

Defining $\omega_{\rm Fit} = \omega_{\rm PRE} (1+\delta)$ for some small $\delta$, we can then write
\begin{equation}
\frac{A^2}{B^2} = \frac{\omega_{\rm PRE}^2 (1+\delta)^2 - \omega_{\rm MILES}^2}{\omega_{\rm PRE}^2 - \omega_{\rm MILES}^2}
\label{ab.eqn}
\end{equation}
Dropping terms $\mathcal{O} (\delta^2)$, we can solve Equation \ref{ab.eqn} for $\delta$ and find
\begin{equation}
\delta = \frac{\omega_{\rm PRE}^2 - \omega_{\rm MILES}^2}{2 \omega_{\rm PRE}^2} \left( \frac{A^2}{B^2} -1 \right)
\end{equation}

\begin{figure}
\begin{center}
\includegraphics[width=\columnwidth]{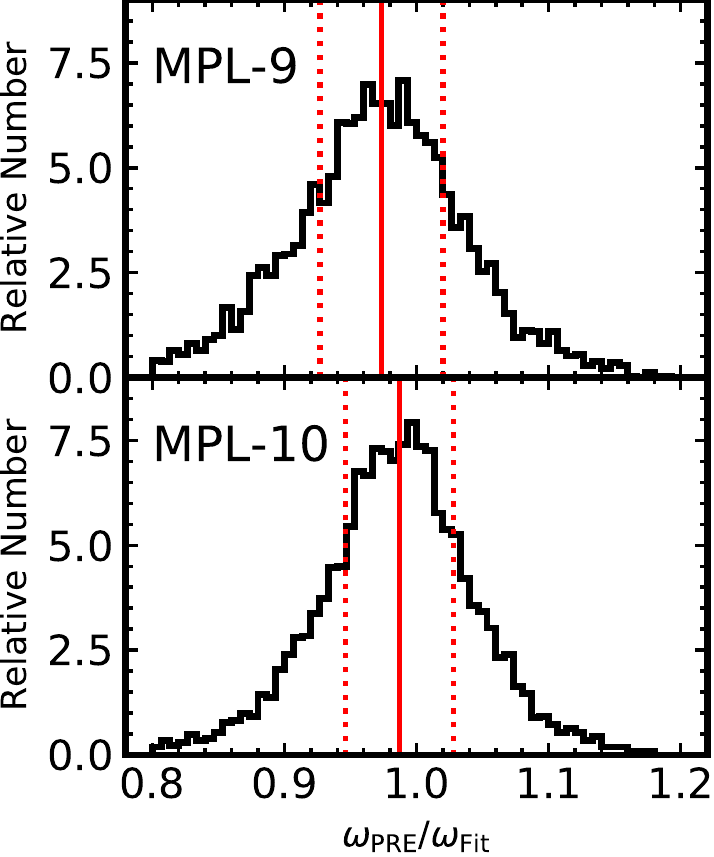}
\end{center}
\caption{Ratio of the MaNGA LSF estimated from the pipeline ($\omega_{\rm PRE}$) vs that derived from full spectral fitting with a series of MILES-HC templates ($\omega_{\rm Fit}$) for  5,000 MaSTAR stellar spectra in MPL-9 and MPL-10.  In each panel the solid red line
indicates the sigma-clipped mean of the distribution, while the dotted red lines indicate the $1\sigma$ width of the distribution.
}
\label{mastar_compare.fig}
\end{figure}

In Fig.~\ref{mastar_compare.fig}, we show the distribution of $\omega_{\rm PRE}/\omega_{\rm Fit} = \frac{1}{1+\delta}$ for the
5,000 MaSTAR spectra in our sample.  For MPL-9, we find that the distribution has a sigma-clipped mean of 0.97 and $1\sigma$ width of 0.05; i.e., suggesting that the the MPL-9 LSF is underestimated by about 3\%.  In contrast, for MPL-10 the distribution has a sigma clipped mean of 0.99 and a $1\sigma$ width of 0.04, indicating both that the scatter of the distribution has decreased and the overall agreement 
 between the pipeline and MILES estimates has improved to within 1\%.\footnote{\citet[][Fig.~17]{westfall19} shows the results of a similar test, finding $A/B = 1.00$ for MPL-9, corresponding to
  $\omega_{\rm MPL-9}/\omega_{\rm Fit} = 1.00$. However, our previous analysis mistakenly used the post-pixel LSF estimate $\omega_{\rm POST}$ instead
 of $\omega_{\rm PRE}$ as should have been applied for a spectral convolution kernel, which approximately canceled out the error in the MPL-9 LSF (Figure \ref{comparedr.fig}).}
 This result is broadly consistent with Figure \ref{comparedr.fig}, which found that the MPL-9
 LSF estimate was $\sim 2-3$\% narrower than the MPL-10
 estimate in the wavelength range covered by the MILES library.
 
We note a few important caveats to this result however.  First, the stellar absorption line kinematics are the product of a convolution across a wide range of wavelengths across which the MaNGA LSF varies substantially.  As such, the actual accuracy of the pipeline LSF at a given wavelength may be better or worse than 1\%, depending on the effective weight of that wavelength range in driving the stellar population model fit. Likewise, our result is predicated upon the assumption that the MILES spectral resolution itself has $< 1$\% systematic error \citep[statistical uncertainties
are known to be $\sim 3$\%, see][]{falconbarroso11},
and that there are negligible systematic differences due to template mismatch between MILES-HC and the MaSTAR sample. 

\subsection{Comparison to the DiskMass Survey}
\label{diskmass.sec}


Perhaps the strongest possible test of the MaNGA LSF model for emission-line kinematics is to compare the derived galaxy-resolved velocity dispersion profile $\sigma_{\Ha}$ in the face of LSF variations, beam smearing, and other factors against prior observations of the same galaxies from higher resolution IFU observations.  We therefore compared the MaNGA data
against \Ha\ observations\footnote{Although the DiskMass survey observed \othree\ as well, MaNGA does not detect \othree\ from DiskMass galaxies at a sufficiently high SNR to enable
a robust comparison.} from the 
DiskMass survey \citep{bershady10a,bershady10b,westfall11,westfall14,martin13} which used the SparsePak IFU \citep{bershady04,bershady05} on the 3.5m WIYN telescope to obtain $R \sim 10,000$ ($\sigma_{\rm inst}=12.7$ \kms) fiber spectroscopy of 176 spiral galaxies oriented nearly face-on to the line of sight. SparsePak fibers have 4.7 arcsec diameters. As of MPL-10, MaNGA has observed seven galaxies in common with DiskMass (see Table \ref{diskmass.table}), which can be identified via targeting bit $2^{16}$ in the ancillary target flag MANGA\_TARGET3.

\begin{deluxetable}{lllll}
\tablecolumns{5}
\tablewidth{0pc}
\tabletypesize{\scriptsize}
\tablecaption{MaNGA DiskMass Overlap Sample}
\tablehead{
\colhead{Plate-IFU} &  \colhead{Name} & \colhead{Redshift\tablenotemark{a}} & \colhead{Inclination\tablenotemark{a}} & \colhead{Stellar Mass\tablenotemark{a}}\\
& & & & \colhead{(log($M_{\ast}/M_{\odot}$)}}
\startdata
8566-12705 & UGC 3997 & 0.0198 & $32^{\circ}$ &  10.1 \\
8567-12701 & UGC 4107 & 0.0117 & $24^{\circ}$ & 10.4 \\
8569-12705\tablenotemark{b} &  UGC 463 & 0.0148 & $28^{\circ}$ & 10.8 \\
8570-9101\tablenotemark{b} & UGC 1087 & 0.0152 & $16^{\circ}$ & 10.3 \\
8939-12704 & UGC 4368 & 0.0129 & $36^{\circ}$ & 10.5 \\
10494-12705 & UGC 4380 & 0.0249 & $16^{\circ}$ & 10.8 \\
10510-12704 & UGC 6918 & 0.0037 & $30^{\circ}$ & 10.0
\enddata
\tablenotetext{a}{Derived from the NASA Sloan Atlas \citep{blanton11} assuming $h=0.7$ and a \citet{chabrier03} IMF.}
\tablenotetext{b}{Not observed prior to MPL-10.}
\label{diskmass.table}
\end{deluxetable}

\begin{figure*}
\begin{center}
\includegraphics[width=\textwidth]{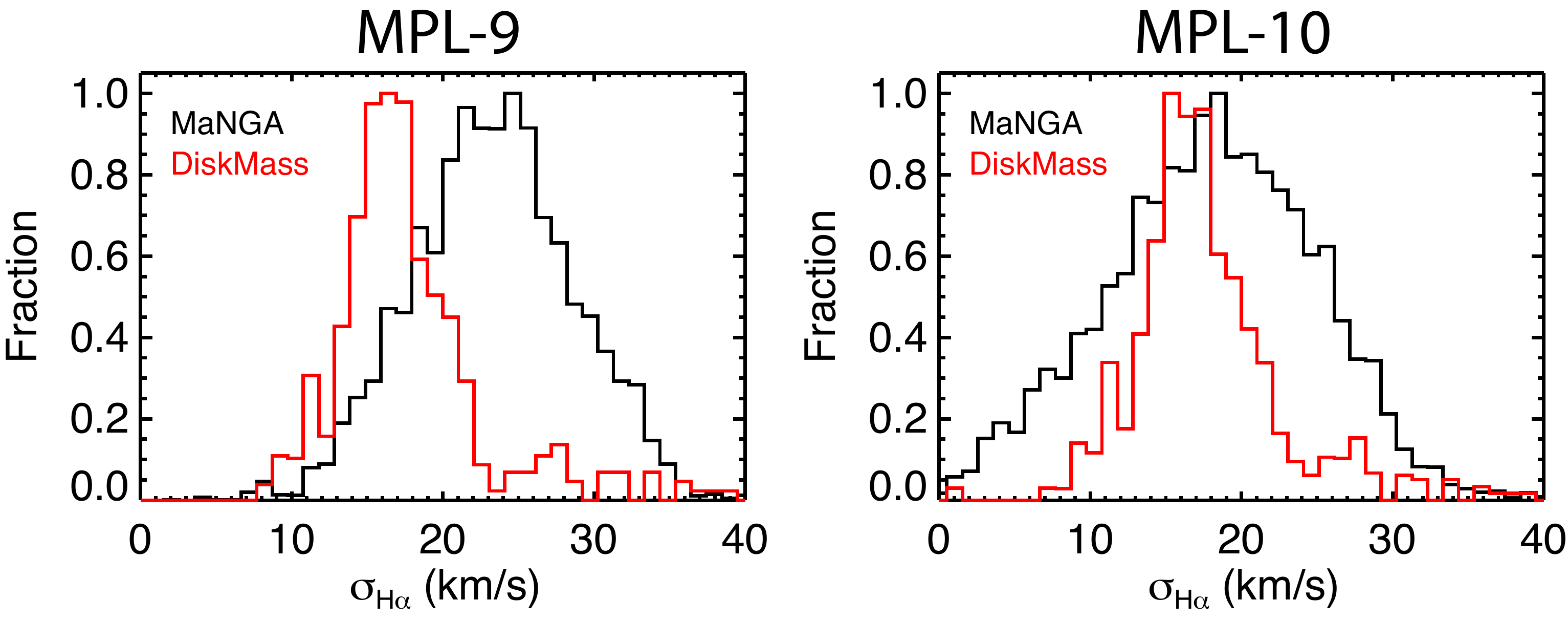}
\end{center}
\caption{Distribution of \Ha\ gas-phase velocity dispersions in the radius range $4 < r < 15$ arcsec for the seven galaxies observed in common with the DiskMass survey.  The peak of the MPL-9 
distribution is offset from the DiskMass distribution by about 7 \kms, while the peak of the MPL-10 distribution matches to within 1.5 \kms.  Note that the DiskMass histograms differ slightly between panels as there are fewer galaxies in common with MPL-9 than with MPL-10.
}
\label{diskmass.fig}
\end{figure*}

We compare the MaNGA and DiskMass samples by extracting kinematic data for all good-quality spaxels with SNR $> 50$ in the common region of overlap $4 < r < 15$ arcsec.  This radial cut is designed to exclude the central regions of the galaxies for which beam smearing is most significant and for which at least one galaxy (UGC 4368) exhibits Seyfert-I type AGN contributions to the H$\alpha$ emission. In Figure \ref{diskmass.fig} we plot a histogram\footnote{Strictly, we sum the normalized histograms of $\sigma_{\Ha}$ for each galaxy to ensure that no one galaxy dominates the distribution if it has more high-SNR spaxels than the others.} of the raw DiskMass and MaNGA measurements 
(i.e., uncorrected for beam smearing)
for both MPL-9 and MPL-10. 
As illustrated in the left-hand panel, MPL-9 is appreciably biased with respect to the DiskMass data, peaking at 23.3 \kms\ instead of 16.7 \kms\ indicative of a 2.9\% systematic underestimate in the LSF. In contrast, the MPL-10 histogram peaks at 18.2 \kms\, matching the DiskMass data to within 0.6\% systematic error in the LSF around \Ha.
If we account for the expected positivity bias in the MaNGA observations from Figure \ref{uncertainty.fig}, this agreement improves further to about 17.4 \kms, or about 0.3\% systematic error in the LSF.
Similarly, this level of agreement is largely insensitive to 
whether or not we apply a beam smearing correction to the MaNGA data; 
following the method described in \S \ref{beamsmear.sec} the corrected MPL-10 data
matches the DiskMass observations to within 0.3\%, and we obtain a comparable result if we instead estimate the beam smearing correction using 
a Bayesian forward model of a disk-like rotation curve.


We can also use the relative widths of the DiskMass and MaNGA MPL-10 distributions to assess the statistical error in  individual estimates of the LSF. Assuming that the DiskMass histogram represents the true astrophysical range of $\sigma_{\Ha}$\footnote{Accounting for the median 1 \kms\ uncertainty in the DiskMass measurements has negligible impact on our results.}, we  construct a Monte Carlo simulation with line widths values drawn from the DiskMass distribution and convolve them with our median LSF of 68.5 \kms\ to create a mock set of observations. We then perturb these values by random errors combining the DAP-reported uncertainties in individual measurements (accounting for the plateau at high SNR discussed in \S \ref{performance.sec}) and some statistical uncertainty in the LSF.  After subtracting the LSF from these perturbed values in quadrature, we compare the width of the simulated distribution to the MPL-10 observations.  This exercise suggests that the statistical uncertainty in the LSF around \Ha\ for a given spaxel is about 2\%, corresponding to 1.4 \kms\ (i.e., comparable to the uncertainty in the measured line widths at high SNR).

\subsection{Comparison to the SAMI Survey}
\label{sami.sec}

As a final consistency check, we additionally compare the MaNGA data
against similar observations obtained by the SAMI survey \citep{bryant15}
using the Sydney-AAO integral field spectrograph \citep{croom12} on the Anglo-Australian
Telescope.  While SAMI has a similar spectral resolution to MaNGA in its blue arm,
around \Ha\ SAMI has a spectral resolution $R \sim 4300$ delivering a $1\sigma$
LSF $\omega_{\rm SAMI} = 30$ \kms.

In MPL-10 we find that there are 74 targets observed in common between MaNGA and SAMI
DR2 \citep{scott18}, and we select the 23 that have been observed with MaNGA's largest
IFU bundle size (12 of which have significant \Ha\ emission) for comparison.
For each of these 12 galaxies, we extract the DAP kinematic measurements for all 
good-quality spaxels with SNR $> 50$ in the common radial range $3 < r < 7.5$ arcsec, where
the upper boundary is set by the size of the SAMI field coverage.  Similarly, we
extract all of the \Ha\ kinematic measurements for these same galaxies provided by the 
SAMI DR2 public data products, introducing a limiting flux cut
of $2 \times 10^{-17}$ erg s$^{-1}$ cm$^{-2}$ \AA$^{-1}$ spaxel$^{-1}$ which visual
inspection suggests selects a nearly identical range of spaxels for SAMI as the $SNR > 50$
cut does for MaNGA.

\begin{figure*}
\begin{center}
\includegraphics[width=\textwidth]{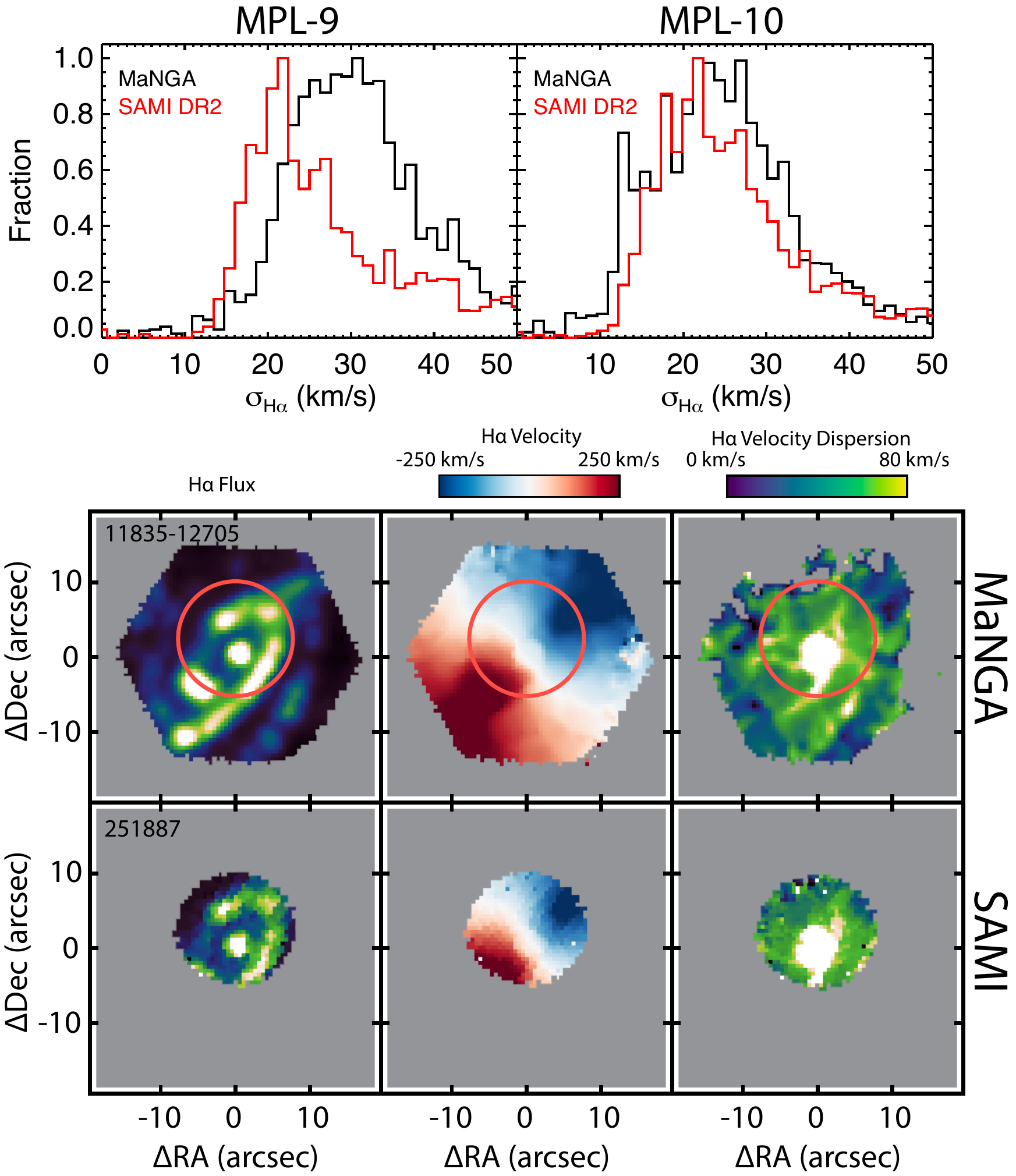}
\end{center}
\caption{Top panels: Distribution of \Ha\ gas-phase velocity dispersions in the radius range
$3 < r < 7.5$ arcsec for twelve galaxies observed in common with the SAMI survey.  The peak of
the MPL-9 distribution is offset from the SAMI DR2 distribution by about 7 \kms, while the peak of the MPL-10 distribution matches to within 0.7 \kms.  SAMI DR2 histograms differ slightly between panels as there are fewer galaxies in common with MPL-9 than with MPL-10.  Bottom panels: Comparison of MaNGA and
SAMI derived \Ha\ flux, velocity, and velocity dispersion maps for example galaxy
11835-12705 (GAMA ID 251887 in SAMI DR2).  The red circles indicate the SAMI field of view.  Note how even small irregularities in the velocity
dispersion maps are seen in both the MaNGA and the SAMI data.
}
\label{sami.fig}
\end{figure*}

As we demonstrate in Figure \ref{sami.fig} (top panels), 
despite the 2.3$x$ higher spectral resolution of SAMI
the MaNGA MPL-10 \Ha\ velocity dispersions are in excellent agreement with the values
provided by SAMI DR2, with the centroids of
the respective histograms matching each other to within 0.7 \kms.  
As expected, the MPL-9 velocity dispersions are in contrast systematically too large by
about 7 \kms, consistent with our comparison against the DiskMass observations.

In addition to the statistical agreement between the MaNGA MPL-10 and SAMI DR2 velocity
dispersions, we note also the excellent agreement in terms of the resolved spatial
structures.  In Figure \ref{sami.fig} (bottom panels), we show an illustrative example
of the \Ha\ flux, velocity, and velocity dispersion maps produced independently by the
two surveys.  The level of agreement between the two is exquisite, with even small and
apparently insignificant features in the dispersion map appearing nearly identically in
both sets of observations.

\section{Summary}
\label{summary.sec}

We have presented a major update to the MaNGA data-reduction pipeline (DRP) that dramatically revises the treatment of the spectral LSF
compared to both previous versions of the MaNGA DRP and the prior SDSS spectroscopic pipeline from which many original MaNGA routines
were derived.  After demonstrating that the LSF can be reliably parameterized as a Gaussian function with $1\sigma$ width
 $\omega$ (Figure \ref{arcline.fig}), we showed in \S \ref{drp.sec} that it is possible to use individual arc lamp exposures in combination with unresolved night-sky emission features to construct a model of the LSF for all of the MaNGA fiber spectra.  These models retain both 
pre-pixellized and post-pixellized measurements (i.e., measurements that either do or do not account
 for the contribution of the top-hat pixel response function to the effective LSF) and are carried through the pipeline
 accounting for various broadening terms to produce three-dimensional LSF cubes corresponding to each of the MaNGA galaxy data cubes.
 
 These LSF data cubes are then used by the MaNGA data-analysis pipeline (DAP; see \S \ref{dap.sec}) to produce kinematic maps that robustly subtract
 the instrumental contribution to the observed spectral line profiles.  An important additional ingredient in such analyses that is not accounted
 for in the default DAP products is beam smearing, wherein the $\sim 2.5$ arcsec FWHM MaNGA spatial PSF in the reconstructed data cubes
 can inflate the apparently line width from unresolved velocity gradients.  As we demonstrated in \S \ref{beamsmear.sec}, 
 although typical beam smearing corrections in the radial range studied here
(i.e., $r > 4$ arcsec) are small they are nonetheless important at the few \kms\ level and must be taken into account by science analyses that
aim to study the cold disk regime around $\sigma_{\Ha} = 10-30$ \kms.

Using Monte Carlo simulations we demonstrated that the MaNGA DAP data products are reliable down to at least $\sigma_{\Ha} = 20$ \kms\
for spaxels with SNR $> 50$ with a typical statistical uncertainty of $4-6$ \kms\ and $< 1$ \kms\ systematic bias (Figure \ref{uncertainty.fig}).  At lower $\sigma_{\Ha}$
and/or lower SNR the data exhibit increasingly large systematic biases towards larger values of $\sigma_{\Ha}$
due to the asymmetric error distribution (i.e., spaxels whose measured line width scatters to below the nominal instrumental
resolution are effectively lost from the sample).
At other wavelengths, we showed in Section \ref{multlines.sec} that 
$\sigma_{\otwo}$, $\sigma_{\Hb}$, $\sigma_{\othree}$, $\sigma_{\ntwo}$, and $\sigma_{\stwo}$
are consistent with $\sigma_{\Ha}$ to within 2\% at $\sigma_{\Ha} > 30$ km s$^{-1}$, 
with possible evidence for systematic variation as a function of ionization energy (Figure \ref{emlines.fig}).

We have confirmed these findings by direct comparison of the MANGA data against higher resolution external data.
Qualitatively, we showed in \S \ref{xshooter.sec} that the pipeline-estimated LSF is consistent to within uncertainty with the LSF derived from
comparing MaSTAR spectra of six bright stars against $R \sim 7000-11,000$ X-Shooter spectra.  Quantitatively, we demonstrated 
in \S \ref{shravan.sec} that the
overall pipeline LSF estimate is consistent at the 1\% level with external assessments using the MILES stellar spectral library to perform full
spectral fitting of 5000 stars drawn at random from the MaSTAR sample.  Further, in \S \ref{diskmass.sec} we demonstrated that the end-to-end
derived MaNGA data products give \Ha\ velocity dispersions peaked around $\sigma_{\Ha} =  18.2$ \kms\ for seven galaxies in common
with the $R \sim 11,000$ DiskMass IFU survey (Figure \ref{diskmass.fig}).  Given the DiskMass result of $\sigma_{\Ha} =  16.7$ \kms, this implies that the MaNGA LSF
in the vicinity of \Ha\ has a systematic uncertainty of $\leq 0.6$\% and a statistical uncertainty of 2\%.  Finally, we showed in \S \ref{sami.sec} that the MaNGA \Ha\ velocity
dispersions are consistent with those derived from $R \sim 4300$ observations from
the SAMI survey, with the distribution of values for a sample of twelve galaxies observed
in common by the two surveys agreeing to within 0.7 \kms.

We therefore conclude that the MaNGA data products provided with internal release MPL-10
are sufficiently well calibrated to allow scientific analysis of the ionized gas velocity dispersions down
to about 20 \kms\ with sufficient care and attention to detail.  Previous public MaNGA data releases
(DR13, DR14, DR15) exhibit few-percent systematic biases in the instrumental LSF however
(see Figure \ref{comparedr.fig}) that will complicate efforts to perform such analyses.
Updated MaNGA products will be released publicly in DR17.

\acknowledgments

DRL appreciates productive discussions with Jeb Bailey, and constructive suggestions by
the anonymous referee.  RY and DL acknowledge support by NSF AST-1715898.  MAB acknowledges NSF Awards AST-1517006 and AST-1814682.

Funding for the Sloan Digital Sky Survey IV has been provided by the Alfred P. Sloan Foundation, the U.S. Department of Energy Office of Science, and the Participating Institutions. SDSS-IV acknowledges
support and resources from the Center for High-Performance Computing at
the University of Utah. The SDSS web site is www.sdss.org.

SDSS-IV is managed by the Astrophysical Research Consortium for the 
Participating Institutions of the SDSS Collaboration including the 
Brazilian Participation Group, the Carnegie Institution for Science, 
Carnegie Mellon University, the Chilean Participation Group, the French Participation Group, Harvard-Smithsonian Center for Astrophysics, 
Instituto de Astrof\'isica de Canarias, The Johns Hopkins University, Kavli Institute for the Physics and Mathematics of the Universe (IPMU) / 
University of Tokyo, the Korean Participation Group, Lawrence Berkeley National Laboratory, 
Leibniz Institut f\"ur Astrophysik Potsdam (AIP),  
Max-Planck-Institut f\"ur Astronomie (MPIA Heidelberg), 
Max-Planck-Institut f\"ur Astrophysik (MPA Garching), 
Max-Planck-Institut f\"ur Extraterrestrische Physik (MPE), 
National Astronomical Observatories of China, New Mexico State University, 
New York University, University of Notre Dame, 
Observat\'ario Nacional / MCTI, The Ohio State University, 
Pennsylvania State University, Shanghai Astronomical Observatory, 
United Kingdom Participation Group,
Universidad Nacional Aut\'onoma de M\'exico, University of Arizona, 
University of Colorado Boulder, University of Oxford, University of Portsmouth, 
University of Utah, University of Virginia, University of Washington, University of Wisconsin, 
Vanderbilt University, and Yale University.

\appendix

\section{Changes to MaNGA fiber bundle metrology}
\label{metrology.sec}

The relative positions of individual fibers within each MaNGA IFU
were measured in the lab to an accuracy of $\sim 0.3$ \micron, corresponding to 5 mas
projected on the sky \citep[][their \S 4.5]{drory15}.  However, the overall scale factor of the lab measurements was less well
calibrated, resulting in a few-percent systematic uncertainty in the size of the IFU as a whole.
With the advantage of years of on-sky observations, it has been possible to use MaNGA observations
to empirically constrain any such systematic scale factors offsets and correct them in
the survey metadata.

As discussed by \citet[][see their \S 8.2]{law16}, the MaNGA DRP includes an `extended astrometry module' (EAM) which compares the galaxy images reconstructed from the IFU data to pre-existing
SDSS broadband imaging photometry in each of the $griz$ bandpasses.  While the EAM is used
automatically to determine any astrometric pointing or rotational offsets in individual exposures
(due, e.g., to inaccuracies in the drilled plate hole locations, telescope pointing/guiding,
and clocking biases from the tension of the IFU fibers within a given cartridge), it is also
possible to adapt it to solve for any global scale factor offsets as well.

\begin{figure}
\begin{center}
\includegraphics[width=\textwidth]{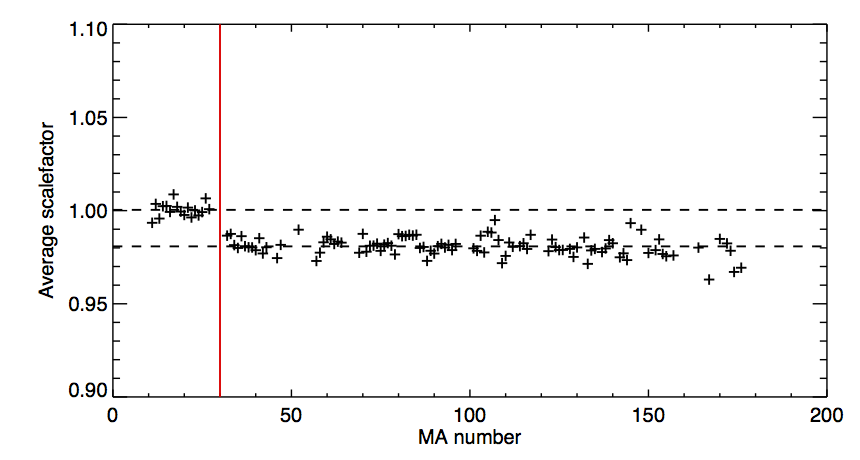}
\epsscale{0.5}
\end{center}
\caption{Fiber bundle size scaling factor derived from comparing reconstructed MaNGA IFU
data against SDSS broadband imaging as a function of the IFU harness identifier.  Values shown
represent averages over all galaxies observed with a given IFU.  Note the 2\% offset between
empirical measurement and lab-derived scaling factors for harnesses in the MaNGA production
run (MA part numbers greater than 30) compared to the commissioning harnesses.  Gaps in the 
plotted points correspond to 7-fiber minibundles for which no scaling information is available.
}
\label{metrology.fig}
\end{figure}

In Figure \ref{metrology.fig} we show the results of running the EAM with a global scaling term on all 6779 data cubes in MPL-8 using the original lab-based fiber metrology, and averaging over all
galaxies observed with a given one of the $\sim 100$ MaNGA science IFUs (17 science IFUs in
each of six carts, with some additional spares swapped in over the lifetime of the survey).  Although the optimal scale factors derived from individual galaxies can by noisy (particularly
for relatively featureless galaxies), the average over many tens of galaxies per fiber bundle
is extremely well behaved and shows that while the initial 30 IFUs built for MaNGA commissioning 
were correct to better than 1\% the remaining IFUs built during production had 
lab-measured scale factors that were systematically too large by 2\%.

In v2.5.3 of the MaNGA DRP we corrected the fiber bundle metrology for this 2\% scale error,
along with an additional 0.5\% in v2.7.1 based on an improved analysis permitted by the
increasing number of galaxy observations.  After applying these corrections, all scale factors
derived by the EAM are consistent with unity to within 0.3\%.

In practice, the impact of these changes between DR15 (v2.4.3) and MPL-10 are minimal
and too small to detect for individual galaxies since a 2\% scale factor change corresponds
to a 0.5 arcsec astrometric shift at the edges of the largest MaNGA fiber bundles.
However, since the metrology of the calibration minibundles also changed there was a corresponding
change in the derived flux calibration.  Since the bundles effectively got slightly smaller the derived PSF shrank, corresponding to reduced throughput at fixed recovered values, causing a correction that produces data cubes whose fluxes are systematically brighter. In combination with
the change from the \citet{od94} to \citet{fitz99} extinction curves, the typical galaxy thus
became brighter by about 3\% in v2.5.3 compared to v2.4.3.

\section{Mitigation of the r1 `blowtorch' artifact}
\label{blowtorch.sec}

Starting in Summer 2019, the r1 detector (one of the four BOSS CCDs) developed a persistent electronic
artifact dubbed the `blowtorch' in which a region of extremely bright pixels produced
a widespread glow that contaminated the lower-left region of the detector.  Although 
multiple efforts were made to identify and fix the physical cause of the artifact, none
of these efforts were wholly successful and the MaNGA DRP therefore had to be modified
to satisfactorily model and subtract this artifact from much of the final year of
observational data.

\begin{figure*}
\begin{center}
\includegraphics[width=\textwidth]{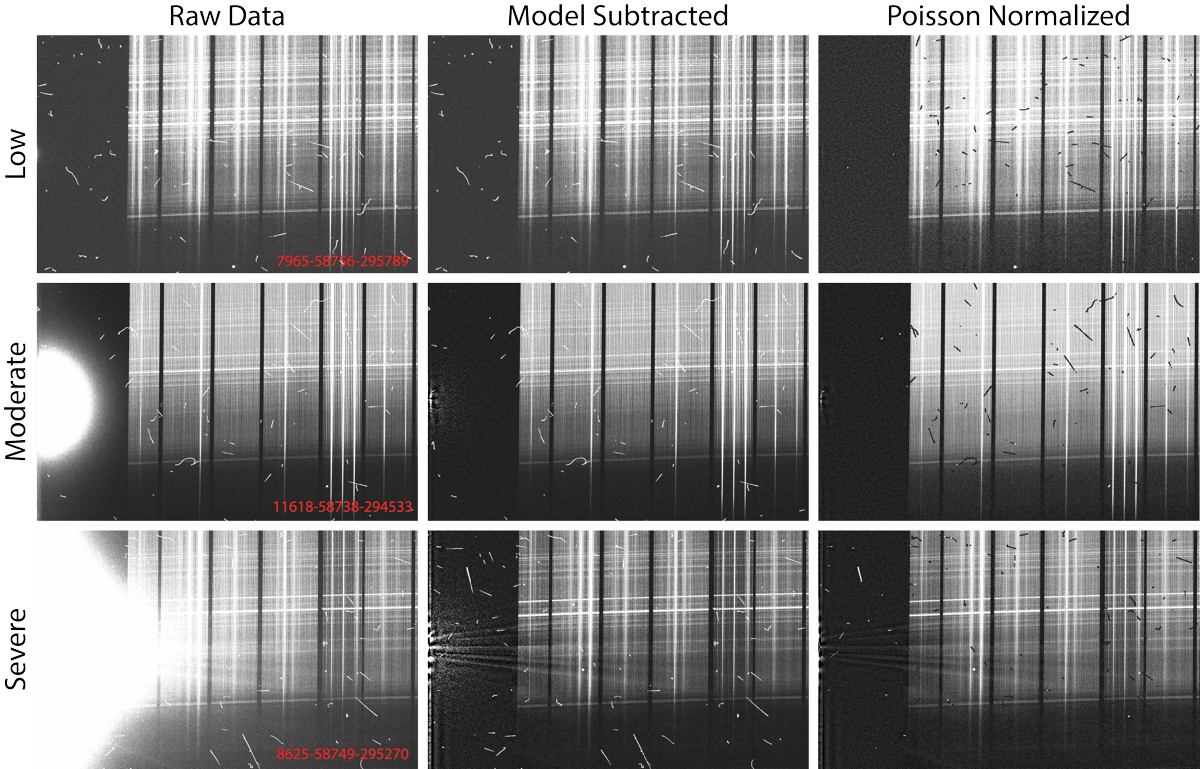}
\epsscale{0.5}
\end{center}
\caption{Left column: Section of the raw detector images (after processing to remove quadrant-dependent
bias, pixel flatfield, and overscan) for the BOSS r1 detector during the 2019-2020 survey
year illustrating the presence of an electronic artifact dubbed the `blowtorch'.  The strength of this artifact is variable, and examples are show for a low (3 e- pixel$^{-1}$),
moderate (1022 e- pixel$^{-1}$), and severe (9831 e- pixel$^{-1}$) case.  Red inset text gives the plate, MJD, and exposure
number of the example frame.  Middle column: Residual detector images after subtraction
of a spline-based model for the artifact.  Right column: Residual detector images after
normalization by the expected poisson noise based on the observed counts in each pixel.
White and/or black streaks in each image are due to cosmic rays, which are automatically
detected and masked by the DRP.
}
\label{blowtorch.fig}
\end{figure*}

As illustrated in Figure \ref{blowtorch.fig} (left column), the strength of the artifact was variable
with time ranging from low (median signal $\leq$ 30 e- pixel$^{-1}$ in the peak affected region),
to moderate (median signal $\leq$ 1500 e- pixel$^{-1}$), to severe (up to about 9000 e- pixel$^{-1}$).  Of the 1878 MaNGA science exposures from the final year of the survey,
157/1471/250 fall into each of these three categories respectively.

As of DRP v3.0.1, for all frames in which the artifact is greater than 30 e- pixel$^{-1}$
in strength the DRP creates a model for the blowtorch component by first masking out all
pixels near the fiber traces and then going row-by-row through the data fitting
a cubic spline function to the unmasked data points.  This spline is constrained to have
more closely-spaced breakpoints in the region nearest to the artifact and widely spaced
breakpoints at larger distance to avoid unphysical structure in the model far from the
artifact.  These row-by-row models are then fit with a second spline model running
column-by-column in order to enforce smoothness of the final model in both detector
dimensions.

Figure \ref{blowtorch.fig} (middle column) demonstrates that the residual detector image
after subtraction of the spline model is relatively clean.  Indeed, for moderate severity
artifacts the poisson-normalized image 
(i.e., the residual count image divided by the shot noise) in Figure \ref{blowtorch.fig} (right column) shows that the subtraction leaves no artifacts in the region of the fiber
traces other than a slightly higher noise.
The DRP flags all data cubes with science exposures in this moderate category with the `BLOWTORCH' flag in the {\sc drp3qual} maskbit, although it is not expected to appreciably
impact the science data quality.

Even in the most severe cases (Figure \ref{blowtorch.fig}, bottom row) the subtraction does
an excellent job, although the SNR degradation is significant (a factor $\sim 2-3$)
and there are residual spokes extending into the science data that were unable to be
modelled.  These spokes will manifest as unidentified emission features in the 6100-6300 \AA\ range for the 3 IFUs on the left edge of the detector, and the DRP therefore flags
all data cubes with science exposures in this category with the `SEVEREBT' and `UNUSUAL'
flags in {\sc drp3qual}.  These data cubes are not included in the count of high quality
galaxy data cubes, although they will nonetheless be acceptable for the vast majority
of science use cases.

\end{document}